\input harvmac

\input amssym
\input epsf

% FONTS

% fraktur
\newfam\frakfam
\font\teneufm=eufm10
\font\seveneufm=eufm7
\font\fiveeufm=eufm5
\textfont\frakfam=\teneufm
\scriptfont\frakfam=\seveneufm
\scriptscriptfont\frakfam=\fiveeufm

% black board bold

% double stroke math

\newfam\dsromfam
\font\tendsrom=dsrom10
\textfont\dsromfam=\tendsrom
\def\ds{\fam\dsromfam \tendsrom}

% bold math italics

\newfam\mbffam
\font\tenmbf=cmmib10
\font\sevenmbf=cmmib7
\font\fivembf=cmmib5
\textfont\mbffam=\tenmbf
\scriptfont\mbffam=\sevenmbf
\scriptscriptfont\mbffam=\fivembf

% bold math cal

\newfam\mbfcalfam
\font\tenmbfcal=cmbsy10
\font\sevenmbfcal=cmbsy7
\font\fivembfcal=cmbsy5
\textfont\mbfcalfam=\tenmbfcal
\scriptfont\mbfcalfam=\sevenmbfcal
\scriptscriptfont\mbfcalfam=\fivembfcal

% math script

\newfam\mscrfam
\font\tenmscr=rsfs10
\font\sevenmscr=rsfs7
\font\fivemscr=rsfs5
\textfont\mscrfam=\tenmscr
\scriptfont\mscrfam=\sevenmscr
\scriptscriptfont\mscrfam=\fivemscr

% MACROS

% bras, kets, ...

\def\vev#1{\left\langle #1\right\rangle}

% tilde, hat, bar, ...

\def\hat{\widehat}

\def\bar{\overline}
\def\b{\bar}
\def\bsq#1{{{\b{#1}}^{\lower 2.5pt\hbox{$\scriptstyle 2$}}}}
\def\bexp#1#2{{{\b{#1}}^{\lower 2.5pt\hbox{$\scriptstyle #2$}}}}
\def\dotexp#1#2{{{#1}^{\lower 2.5pt\hbox{$\scriptstyle #2$}}}}

\def\hh{{\hat h}}
\def\hc{{\hat c}}
\def\hlambda{{\hat \lambda}}

% basic math
\def\IL{\relax{\rm I\kern-.18em L}}
\def\IH{\relax{\rm I\kern-.18em H}}
\def\IR{\relax{\rm I\kern-.18em R}}
\def\IC{\relax{\rm I\kern-0.54 em C}}

\def\rt2{\sqrt{2}}
\def\half {{1 \over 2}}

% bold greek characters

\font\tenbifull=cmmib10
\font\tenbimed=cmmib7
\font\tenbismall=cmmib5
\textfont9=\tenbifull \scriptfont9=\tenbimed
\scriptscriptfont9=\tenbismall

\mathchardef\bbGamma="7000
\mathchardef\bbDelta="7001
\mathchardef\bbPhi="7002
\mathchardef\bbAlpha="7003
\mathchardef\bbXi="7004
\mathchardef\bbPi="7005
\mathchardef\bbSigma="7006
\mathchardef\bbUpsilon="7007
\mathchardef\bbTheta="7008
\mathchardef\bbPsi="7009
\mathchardef\bbOmega="700A
\mathchardef\bbalpha="710B
\mathchardef\bbbeta="710C
\mathchardef\bbgamma="710D
\mathchardef\bbdelta="710E
\mathchardef\bbepsilon="710F
\mathchardef\bbzeta="7110
\mathchardef\bbeta="7111
\mathchardef\bbtheta="7112
\mathchardef\bbkappa="7114
\mathchardef\bblambda="7115
\mathchardef\bbmu="7116
\mathchardef\bbnu="7117
\mathchardef\bbxi="7118
\mathchardef\bbpi="7119
\mathchardef\bbrho="711A
\mathchardef\bbsigma="711B
\mathchardef\bbtau="711C
\mathchardef\bbupsilon="711D
\mathchardef\bbphi="711E
\mathchardef\bbchi="711F
\mathchardef\bbpsi="7120
\mathchardef\bbomega="7121
\mathchardef\bbvarepsilon="7122
\mathchardef\bbvartheta="7123
\mathchardef\bbvarpi="7124
\mathchardef\bbvarrho="7125
\mathchardef\bbvarsigma="7126
\mathchardef\bbvarphi="7127

\def\IL{\relax{\rm I\kern-.18em L}}
\def\IH{\relax{\rm I\kern-.18em H}}
\def\IR{\relax{\rm I\kern-.18em R}}
\def\IC{\relax{\rm I\kern-0.54 em C}}

% dotted spinor indices

% bared indices

% bared spinors

% capital cal letters

\def\CO{{\cal O}}

% double stroke symbols: unit matrix, reals, complex, quaternions, integers, natural numbers

\def\1{{\ds 1}}

% miscellaneous objects

\noblackbox

\def\unit{\relax{\rm 1\kern-.26em I}}
\def\nada{\relax{\rm 0\kern-.30em l}}

%\def\Omega{\rho,\sigma,\nu  }

%% MACROS
\noblackbox
\def\IL{\relax{\rm I\kern-.18em L}}
\def\IH{\relax{\rm I\kern-.18em H}}
\def\IR{\relax{\rm I\kern-.18em R}}
\def\IC{\relax\hbox{$\inbar\kern-.3em{\rm C}$}}
\def\IZ{\relax\ifmmode\mathchoice
{\hbox{\cmss Z\kern-.4em Z}}{\hbox{\cmss Z\kern-.4em Z}} {\lower.9pt\hbox{\cmsss Z\kern-.4em Z}}
{\lower1.2pt\hbox{\cmsss Z\kern-.4em Z}}\else{\cmss Z\kern-.4em Z}\fi}

\def\partialslash{\not{\hbox{\kern-2pt $\partial$}}}

\def\CO {{\cal O}}

%% MORE MACROS

\def\CO {{\cal O}}

\font\manual=manfnt \def\dbend{\lower3.5pt\hbox{\manual\char127}}

\def\IZ{\relax\ifmmode\mathchoice
{\hbox{\cmss Z\kern-.4em Z}}{\hbox{\cmss Z\kern-.4em Z}} {\lower.9pt\hbox{\cmsss Z\kern-.4em Z}}
{\lower1.2pt\hbox{\cmsss Z\kern-.4em Z}}\else{\cmss Z\kern-.4em Z}\fi}
\def\half {{1\over 2}}

\def\bar{\overline}

\def\rt2{\sqrt{2}}
\def\irt2{{1\over\sqrt{2}}}

\def\hat{\widehat}
%  \slashchar puts a slash through a character to represent contraction
%  with Dirac matrices. Use \not instead for negation of relations, and use
%  \hbar for hbar.
\def\slashchar#1{\setbox0=\hbox{$#1$}           % set a box for #1
   \dimen0=\wd0                                 % and get its size
   \setbox1=\hbox{/} \dimen1=\wd1               % get size of /
   \ifdim\dimen0>\dimen1                        % #1 is bigger
      \rlap{\hbox to \dimen0{\hfil/\hfil}}      % so center / in box
      #1                                        % and print #1
   \else                                        % / is bigger
      \rlap{\hbox to \dimen1{\hfil$#1$\hfil}}   % so center #1
      /                                         % and print /
   \fi}

\def\foursqr#1#2{{\vcenter{\vbox{
    \hrule height.#2pt
    \hbox{\vrule width.#2pt height#1pt \kern#1pt
    \vrule width.#2pt}
    \hrule height.#2pt
    \hrule height.#2pt
    \hbox{\vrule width.#2pt height#1pt \kern#1pt
    \vrule width.#2pt}
    \hrule height.#2pt
        \hrule height.#2pt
    \hbox{\vrule width.#2pt height#1pt \kern#1pt
    \vrule width.#2pt}
    \hrule height.#2pt
        \hrule height.#2pt
    \hbox{\vrule width.#2pt height#1pt \kern#1pt
    \vrule width.#2pt}
    \hrule height.#2pt}}}}
\def\psqr#1#2{{\vcenter{\vbox{\hrule height.#2pt
    \hbox{\vrule width.#2pt height#1pt \kern#1pt
    \vrule width.#2pt}
    \hrule height.#2pt \hrule height.#2pt
    \hbox{\vrule width.#2pt height#1pt \kern#1pt
    \vrule width.#2pt}
    \hrule height.#2pt}}}}
\def\sqr#1#2{{\vcenter{\vbox{\hrule height.#2pt
    \hbox{\vrule width.#2pt height#1pt \kern#1pt
    \vrule width.#2pt}
    \hrule height.#2pt}}}}

\def\figin{\epsfcheck\figin}\def\figins{\epsfcheck\figins}
\def\epsfcheck{\ifx\epsfbox\UnDeFiNeD
\message{(NO epsf.tex, FIGURES WILL BE IGNORED)}
\gdef\figin##1{\vskip2in}\gdef\figins##1{\hskip.5in}% blank space instead
\else\message{(FIGURES WILL BE INCLUDED)}%
\gdef\figin##1{##1}\gdef\figins##1{##1}\fi}
\def\DefWarn#1{}
\def\figinsert{\goodbreak\midinsert}
\def\ifig#1#2#3{\DefWarn#1\xdef#1{fig.~\the\figno}
\writedef{#1\leftbracket fig.\noexpand~\the\figno}%
\figinsert\figin{\centerline{#3}}\medskip\centerline{\vbox{\baselineskip12pt \advance\hsize by
-1truein\noindent\footnotefont{\bf Fig.~\the\figno:\ } \it#2}}
\bigskip\endinsert\global\advance\figno by1}

% REFERENCES

%\GubserVV
\lref\GubserVV{
  S.~S.~Gubser and I.~R.~Klebanov,
  ``A Universal result on central charges in the presence of double trace deformations,''
Nucl.\ Phys.\ B {\bf 656}, 23 (2003).
[hep-th/0212138].
%%CITATION = hep-th/0212138%%
}

%\ZamolodchikovGT
\lref\ZamolodchikovGT{
  A.~B.~Zamolodchikov,
  ``Irreversibility of the Flux of the Renormalization Group in a 2D Field Theory,''
JETP Lett.\  {\bf 43}, 730 (1986), [Pisma Zh.\ Eksp.\ Teor.\ Fiz.\  {\bf 43}, 565 (1986)].
}

%\HartnollRZA
\lref\HartnollRZA{
  S.~A.~Hartnoll, D.~M.~Ramirez and J.~E.~Santos,
  ``Thermal conductivity at a disordered quantum critical point,''
[arXiv:1508.04435 [hep-th]].
%%CITATION = arXiv:1508.04435%%
}

%\OsbornGM
\lref\OsbornGM{
  H.~Osborn,
  ``Weyl consistency conditions and a local renormalization group equation for general renormalizable field theories,''
Nucl.\ Phys.\ B {\bf 363}, 486 (1991).
%%CITATION = DAMTP-91-1%%
}

%\MaldacenaRE
\lref\Mal{
  J.~M.~Maldacena,
  ``The Large N limit of superconformal field theories and supergravity,''
Int.\ J.\ Theor.\ Phys.\  {\bf 38}, 1113 (1999), [Adv.\ Theor.\ Math.\ Phys.\  {\bf 2}, 231 (1998)].
[hep-th/9711200].
%%CITATION = HUTP-97-A097%%
}

%\GubserBC
\lref\Gubs{
  S.~S.~Gubser, I.~R.~Klebanov and A.~M.~Polyakov,
  ``Gauge theory correlators from noncritical string theory,''
Phys.\ Lett.\ B {\bf 428}, 105 (1998).
[hep-th/9802109].
%%CITATION = hep-th/9802109%%
}

%\WittenQJ
\lref\Wit{
  E.~Witten,
  ``Anti-de Sitter space and holography,''
Adv.\ Theor.\ Math.\ Phys.\  {\bf 2}, 253 (1998).
[hep-th/9802150].
%%CITATION = hep-th/9802150%%
}

%\HartnollCUA
\lref\HartnollCUA{
  S.~A.~Hartnoll and J.~E.~Santos,
  ``Disordered horizons: Holography of randomly disordered fixed points,''
Phys. Rev.\ Lett.\  {\bf 112}, 231601 (2014).
[arXiv:1402.0872 [hep-th]].
%%CITATION = arXiv:1402.0872%%
}

%\HartnollFAA
\lref\HartnollFAA{
  S.~A.~Hartnoll, D.~M.~Ramirez and J.~E.~Santos,
  ``Emergent scale invariance of disordered horizons,''
[arXiv:1504.03324 [hep-th]].
%%CITATION = arXiv:1504.03324%%
}

%\WittenUA
\lref\WittenUA{
  E.~Witten,
  ``Multitrace operators, boundary conditions, and AdS / CFT correspondence,''
[hep-th/0112258].
%%CITATION = hep-th/0112258%%
}

%\CardyCWA
\lref\CardyCWA{
  J.~L.~Cardy,
  ``Is There a c Theorem in Four-Dimensions?,''
Phys.\ Lett.\ B {\bf 215}, 749 (1988).
}

%\JackEB
\lref\JackEB{
  I.~Jack and H.~Osborn,
  ``Analogs for the $c$ Theorem for Four-dimensional Renormalizable Field Theories,''
Nucl.\ Phys.\ B {\bf 343}, 647 (1990).
%%CITATION = DAMTP-90-02%%
}

%\CaselleCSA
\lref\CaselleCSA{
  M.~Caselle, G.~Costagliola and N.~Magnoli,
  ``Numerical determination of the operator-product-expansion coefficients in the 3D Ising model from off-critical correlators,''
Phys.\ Rev.\ D {\bf 91}, no. 6, 061901 (2015).
[arXiv:1501.04065 [hep-th]].
%%CITATION = arXiv:1501.04065%%
}

%\KomargodskiXV
\lref\KomargodskiXV{
  Z.~Komargodski,
  ``The Constraints of Conformal Symmetry on RG Flows,''
JHEP {\bf 1207}, 069 (2012).
[arXiv:1112.4538 [hep-th]].
%%CITATION = arXiv:1112.4538%%
}

\lref\Gurarie{V.~Gurarie, ``c-Theorem for disordered systems," Nuclear Physics B 546.3 (1999): 765-778. [cond-mat/9808063].}

\lref\StressCardy{J.~Cardy, ``The stress tensor in quenched random systems," Statistical Field Theories 73 (2002): 215-222. [cond-mat/0111031].}

\lref\Hertz{
J.~Hertz, ``Disordered systems," Physica Scripta 1985.T10 (1985): 1.}

%\CardyXT
\lref\CardyXT{
  J.~L.~Cardy,
  ``Scaling and renormalization in statistical physics,''
Cambridge, UK: Univ. Pr. (1996) 238 p. (Cambridge lecture notes in physics: 3).
}

\lref\Belanger{D.~P.~Belanger, ``Experimental characterization of the Ising model in disordered antiferromagnets." Brazilian Journal of Physics 30.4 (2000): 682-692.}

%\ElShowkHT
\lref\ElShowkHT{
  S.~El-Showk, M.~F.~Paulos, D.~Poland, S.~Rychkov, D.~Simmons-Duffin and A.~Vichi,
  ``Solving the 3D Ising Model with the Conformal Bootstrap,''
Phys.\ Rev.\ D {\bf 86}, 025022 (2012).
[arXiv:1203.6064 [hep-th]].
%%CITATION = LPTENS-12-07%%
}

%\FaulknerJY
\lref\FaulknerJY{
  T.~Faulkner, H.~Liu and M.~Rangamani,
  ``Integrating out geometry: Holographic Wilsonian RG and the membrane paradigm,''
JHEP {\bf 1108}, 051 (2011).
[arXiv:1010.4036 [hep-th]].
%%CITATION = arXiv:1010.4036%%
}

%\AharonyZEA
\lref\AharonyZEA{
  O.~Aharony, M.~Berkooz and S.~J.~Rey,
  ``Rigid holography and six-dimensional $ {\cal N}=\left(2,0\right) $ theories on AdS$_{5} \times S^{1}$,''
JHEP {\bf 1503}, 121 (2015).
[arXiv:1501.02904 [hep-th]].
%%CITATION = SNUST-15-01%%
}

\lref\WF{
K.~Wilson and M.~Fisher, ``Critical exponents in 3.99 dimensions," Physical Review Letters 28.4 (1972): 240.
}

%\AdamsRJ
\lref\AdamsRJ{
  A.~Adams and S.~Yaida,
  ``Disordered Holographic Systems I: Functional Renormalization,''
[arXiv:1102.2892 [hep-th]].
%%CITATION = arXiv:1102.2892%%
}

%\AdamsYI
\lref\AdamsYI{
  A.~Adams and S.~Yaida,
  ``Disordered holographic systems: Marginal relevance of imperfection,''
Phys.\ Rev.\ D {\bf 90}, no. 4, 046007 (2014).
[arXiv:1201.6366 [hep-th]].
%%CITATION = MIT-CTP-4344%%
}

%\O'KeeffeAWA
\lref\OKeeffeAWA{
  D.~K.~O'Keeffe and A.~W.~Peet,
  ``Perturbatively charged holographic disorder,''
[arXiv:1504.03288 [hep-th]].
%%CITATION = arXiv:1504.03288%%
}

%\AreanOAA
\lref\AreanOAA{
  D.~Arean, A.~Farahi, L.~A.~Pando Zayas, I.~S.~Landea and A.~Scardicchio,
  ``Holographic p-wave Superconductor with Disorder,''
[arXiv:1407.7526 [hep-th]].
%%CITATION = MCTP-14-16%%
}

%\KlebanovTB
\lref\KlebanovTB{
  I.~R.~Klebanov and E.~Witten,
  ``AdS / CFT correspondence and symmetry breaking,''
Nucl.\ Phys.\ B {\bf 556}, 89 (1999).
[hep-th/9905104].
%%CITATION = hep-th/9905104%%
}

\lref\IM{
Y.~Imry and S-K.~Ma,
  ``Random-field instability of the ordered state of continuous symmetry," Phys.\ Rev.\ Letters\ {\bf 35}, 1399 (1975).
}

\lref\AYS{
A.~Aharony, Y.~Imry, and S-K.~Ma, ``Lowering of dimensionality in phase transitions with random fields," Phys.\ Rev.\  Lett.\ {\bf 37}, 1364 (1976).}

\lref\Grinstein{G.~Grinstein, ``Ferromagnetic phase transition in random field: the breakdown of scaling laws," Phys.\  Rev.\  Lett.\ {\bf 37}, 944 (1976).}

\lref\Young{A.~P.~Young, ``On the lowering of dimenionality in phase transitions with random fields,'' J.~Phys.~C\ {\bf 10}, L257 (1977).}

%\ParisiKA
\lref\ParisiKA{
  G.~Parisi and N.~Sourlas,
  ``Random Magnetic Fields, Supersymmetry and Negative Dimensions,''
Phys.\ Rev.\ Lett.\  {\bf 43}, 744 (1979).
%%CITATION = LPTENS-79-13%%
}

%\El-ShowkDWA
\lref\ElShowkDWA{
  S.~El-Showk, M.~F.~Paulos, D.~Poland, S.~Rychkov, D.~Simmons-Duffin and A.~Vichi,
  ``Solving the 3d Ising Model with the Conformal Bootstrap II. c-Minimization and Precise Critical Exponents,''
J.\ Stat.\ Phys.\  {\bf 157}, 869 (2014).
[arXiv:1403.4545 [hep-th]].
%%CITATION = CERN-PH-TH-2014-038%%
}

\lref\wittenunpub{E. Witten, unpublished.}

%\BardeenPM
\lref\BardeenPM{
  W.~A.~Bardeen and B.~Zumino,
  ``Consistent and Covariant Anomalies in Gauge and Gravitational Theories,''
Nucl.\ Phys.\ B {\bf 244}, 421 (1984).
%%CITATION = FERMILAB-PUB-84-038-T%%
}

\lref\Folk{R.~Folk, Y.~Holovatch, and T.~Yavorskii, ``Critical exponents of a three-dimensional weakly diluted quenched Ising model." Physics-Uspekhi 46.2 (2003): 169-191.}

\lref\MC{
S.~Fan, W.~Xiong, W.~Yuan, and F.~Zhong, ``Critical behavior of a three-dimensional random-bond Ising model using finite-time scaling with extensive Monte Carlo renormalization-group method," Physical Review E 81.5 (2010): 051132.}

%\OsbornGM
\lref\OsbornGM{
  H.~Osborn,
  ``Weyl consistency conditions and a local renormalization group equation for general renormalizable field theories,''
Nucl.\ Phys.\ B {\bf 363}, 486 (1991).
%%CITATION = DAMTP-91-1%%
}

%\CallanSA
\lref\CallanSA{
  C.~G.~Callan, Jr. and J.~A.~Harvey,
  ``Anomalies and Fermion Zero Modes on Strings and Domain Walls,''
Nucl.\ Phys.\ B {\bf 250}, 427 (1985).
%%CITATION = Print-84-0860 (PRINCETON)%%
}

%\ABHarris
\lref\Harris{
  A.~B.~Harris,
  ``Effect of random defects on the critical behaviour of Ising models,''
Journal of Physics C: Solid State Physics 7.9 (1974): 1671.%%CITATION = arXiv:1010.2150%%
}

%\DotsenkoSY
\lref\DotsenkoSY{
  V.~Dotsenko, M.~Picco and P.~Pujol,
  ``Renormalization group calculation of correlation functions for the 2-d random bond Ising and Potts models,''
Nucl.\ Phys.\ B {\bf 455}, 701 (1995).
[hep-th/9501017].
%%CITATION = hep-th/9501017%%
}
%\DotsenkoIM
\lref\DotsenkoIM{
  V.~Dotsenko, M.~Picco and P.~Pujol,
  ``Spin spin critical point correlation functions for the 2-D random bond Ising and Potts models,''
Phys.\ Lett.\ B {\bf 347}, 113 (1995).
[hep-th/9405003].
%%CITATION = hep-th/9405003%%
}

%\ShimadaDM
\lref\ShimadaDM{
  H.~Shimada,
  ``Disordered O(n) Loop Model and Coupled Conformal Field Theories,''
Nucl.\ Phys.\ B {\bf 820}, 707 (2009).
[arXiv:0903.3787 [cond-mat.dis-nn]].
%%CITATION = arXiv:0903.3787%%
}

%\HeemskerkHK
\lref\HeemskerkHK{
  I.~Heemskerk and J.~Polchinski,
  ``Holographic and Wilsonian Renormalization Groups,''
JHEP {\bf 1106}, 031 (2011).
[arXiv:1010.1264 [hep-th]].
%%CITATION = arXiv:1010.1264%%
}

%\Dotsenko
\lref\Dotsenko{
V.~Dotsenko, 
``Introduction to the replica theory of disordered statistical systems,'' Cambridge University Press, 2005.}

%\GliozziYSA
\lref\GliozziYSA{
  F.~Gliozzi,
  ``More constraining conformal bootstrap,''
Phys.\ Rev.\ Lett.\  {\bf 111}, 161602 (2013).
[arXiv:1307.3111].
%%CITATION = arXiv:1307.3111%%
}

%\CardyRQG
\lref\CardyRQG{
  J.~Cardy,
  ``Logarithmic conformal field theories as limits of ordinary CFTs and some physical applications,''
J.\ Phys.\ A {\bf 46}, 494001 (2013).
[arXiv:1302.4279 [cond-mat.stat-mech]].
%%CITATION = arXiv:1302.4279%%
}

%\ShangZW
\lref\ShangZW{
  Y.~Shang,
  ``Correlation functions in the holographic replica method,''
JHEP {\bf 1212}, 120 (2012).
[arXiv:1210.2404 [hep-th]].
%%CITATION = arXiv:1210.2404%%
}

%\FujitaRS
\lref\Taka{
  M.~Fujita, Y.~Hikida, S.~Ryu and T.~Takayanagi,
  ``Disordered Systems and the Replica Method in AdS/CFT,''
JHEP {\bf 0812}, 065 (2008).
[arXiv:0810.5394 [hep-th]].
%%CITATION = arXiv:0810.5394%%
}

\lref\Koma{http://hep.physics.uoc.gr/mideast8/talks/monday/Komargodski.pdf}

%\KiritsisAT
\lref\Niarcho{
  E.~Kiritsis and V.~Niarchos,
  ``Interacting String Multi-verses and Holographic Instabilities of Massive Gravity,''
Nucl.\ Phys.\ B {\bf 812}, 488 (2009).
[arXiv:0808.3410 [hep-th]].
%%CITATION = arXiv:0808.3410%%
}

%\BernardAS
\lref\BernardAS{
  D.~Bernard,
  ``(Perturbed) conformal field theory applied to 2-D disordered systems: An Introduction,''
In *Cargese 1995, Low-dimensional applications of quantum field theory* 19-61.
[hep-th/9509137].
%%CITATION = hep-th/9509137%%
}

%\Cardy
\lref\Cardy{
J.~Cardy, ``Scaling and renormalization in statistical physics,'' Vol. 5. Cambridge university press, 1996.
}

\lref\Ludwig{A.~Ludwig and J.~Cardy. ``Perturbative evaluation of the conformal anomaly at new critical points with applications to random systems." Nuclear Physics B 285 (1987): 687-718.}

%\BanksWHA
\lref\BanksWHA{
  E.~Banks, A.~Donos and J.~P.~Gauntlett,
  ``Thermoelectric DC conductivities and Stokes flows on black hole horizons,''
[arXiv:1507.00234 [hep-th]].
%%CITATION = IMPERIAL-TP-2015-JG-03%%
}

%\ColemanCI
\lref\ColemanCI{
  S.~R.~Coleman,
  ``There are no Goldstone bosons in two-dimensions,''
Commun.\ Math.\ Phys.\  {\bf 31}, 259 (1973).
}

%\ColemanCI
\lref\MW{
D.~Mermin, and H.~Wagner, ``Absence of ferromagnetism or antiferromagnetism in one-or two-dimensional isotropic Heisenberg models," Physical Review Letters 17.22 (1966): 1133.}

%\HeemskerkPN
\lref\HeemskerkPN{
  I.~Heemskerk, J.~Penedones, J.~Polchinski and J.~Sully,
  ``Holography from Conformal Field Theory,''
JHEP {\bf 0910}, 079 (2009).
[arXiv:0907.0151 [hep-th]].
%%CITATION = arXiv:0907.0151%%
}

\def\figcaption#1#2{\DefWarn#1\xdef#1{Figure~\noexpand\hyperref{}{figure}%
{\the\figno}{\the\figno}}\writedef{#1\leftbracket Figure\noexpand~\xfig#1}%
\medskip\centerline{{\footnotefont\bf Figure~\hyperdef\hypernoname{figure}{\the\figno}{\the\figno}:}  #2 \wrlabeL{#1=#1}}%
\global\advance\figno by1}

% PAPER

%\draftmode

%\rightline{}
\rightline{TAUP-2999/15, WIS/04/15-AUG-DPPA}
\vskip-75pt
\Title{
} {\vbox{\centerline{Disorder in Large-$N$ Theories}
%\centerline{     Large N and an Expansion in Heat Capacity }
}}
%\vskip-15pt
\centerline{ Ofer Aharony,$^1$ Zohar Komargodski,$^1$ and Shimon Yankielowicz$^2$}
\vskip15pt

\centerline{ {\it $^1$ Department of Particle Physics and Astrophysics, }}
\centerline{{\it Weizmann Institute of Science, Rehovot
7610001, Israel}}
\centerline{ {\it $^2$ School of Physics and Astronomy, Tel Aviv University, Ramat Aviv 69978, Israel}}
\vskip20pt

\centerline{\bf Abstract}
\noindent

We consider Euclidean Conformal Field Theories perturbed by quenched
disorder, namely by random fluctuations in their couplings. Such
theories are relevant for second-order phase transitions in the
presence of impurities or other forms of disorder. Theories with quenched disorder often flow
to new fixed points of the renormalization group. We begin with disorder in free field theories.
Imry and Ma showed that disordered free fields
can only exist for $d>4$. For $d>4$ we show that
disorder leads to new fixed points which are not scale-invariant. We then move
on to large-$N$ theories (vector models or gauge theories in the `t~Hooft limit). We compute exactly the beta function for the
disorder, and the correlation functions of the disordered theory. We generalize
the results of Imry and Ma by showing that such disordered theories exist
only when disorder couples to operators of dimension $\Delta > d/4$.
Sometimes the disordered fixed points are not scale-invariant, and in other cases they have unconventional dependence on the disorder, including non-trivial effects due to irrelevant operators. Holography maps disorder in conformal theories to stochastic differential equations in a higher dimensional space. We use this dictionary to reproduce our field theory results. We also study the leading
$1/N$ corrections, both by field theory methods and by holography. These corrections
are particularly important when disorder scales with the number of degrees of
freedom.

\Date{September 2015}

\listtoc \writetoc

\newsec{Introduction and Review}

It is of considerable theoretical and experimental interest to consider Quantum Field Theory (QFT) with random couplings. By ``random couplings'' or ``quenched disorder'' we mean that we have some specified probability distribution in the space of couplings constants. In the simplest case one can let the couplings vary independently at every point in Euclidean space (though the ensemble should still consist of  reasonably smooth functions). This probability distribution is independent of the quantum fields, which is why this type of disorder is called ``quenched.'' We imagine first solving the theory for all values of the couplings, and then averaging over the couplings. 

Experimentally, this situation arises (in some cases) for the conformal field theories that describe the critical points of second-order phase transitions, in the presence of disorder. Such disorder is often present, due to impurities or to random external fields.
Upon adding disorder, the phase transition point may be modified (i.e. the critical exponents may change), it can disappear altogether, or it can be unaltered.\foot{We do not study in this paper disordered  theories relevant to quantum phase transitions. In these theories there is an
extra time direction, but the disorder only depends on the spatial directions. We also do not consider thermal phase transitions in quantum systems, which are described by a compact Euclidean time direction. We only consider here the most symmetric case of disorder in Euclidean CFTs; this is relevant to classical phase transitions. We leave the study of the various other cases   by our methods to future work.}

Let us first define the problem (see, for instance,~\refs{\Hertz,\Cardy,\Dotsenko}). Consider a Euclidean QFT in $d$ dimensions, in which all the operators appear in the action with couplings $g^i$ that are allowed to depend on space, $S=S_0+\sum_i\int d^dx g^i(x)\CO_i(x)$. We can then define the generating functional $W[g^i]$ for connected diagrams (i.e. minus the free energy)
\eqn\connected{e^{W[g^i]}\equiv Z[g^i] \equiv \int[{\cal D} \mu]e^{-S_0-\sum_i\int d^dx g^i(x) \CO_i(x)}  ~,}
where $[{\cal D}\mu]$ stands for the path integral measure. In this paper we focus on the case where the theory with $g^i=0$ is a Conformal Field Theory (CFT). Functional derivatives of $W[g^i]$ evaluated at $g^i(x)=0$ give connected correlation functions in the~CFT.

Now, suppose that one of the couplings associated to a scalar operator, $h\equiv g^0$, is random, with a Gaussian probability distribution\foot{The case of more than one coupling being random can be treated similarly, as well as cases where the disorder couples to operators with non-zero spin. Also, one can consider other probability distributions in the space of couplings.} 
\eqn\prob{\overline{h(x)}=0~,\qquad  \overline{h(x)h(y)}=c^2\delta^{(d)}(x-y)~,}
where we denote by $\overline{X}$ the average of $X$ over the disorder.
The disorder average of higher products of $h(x_i)$ is given by a sum over all the two-point functions as in Wick's theorem. 

We can then define the disordered free-energy $W_D$ as an average over the ordinary free-energy  
\eqn\disorder{W_D[g^1,g^2,...]\equiv\int {\cal D} h\  W[h,g^1,g^2...] e^{-{1\over 2c^2}\int d^dx h^2(x)}~,}
and the disorder-averaged vacuum-normalized correlation functions are given by 
\eqn\disorderi{\overline{\langle \CO_1(x_1)...\CO_n(x_n) \rangle} \equiv\int {\cal D} h\  e^{-{1\over 2c^2}\int d^dx h^2(x)}\langle \CO_1(x_1)...\CO_n(x_n) \rangle_h~,}
where $\langle \CO_1(x_1)...\CO_n(x_n) \rangle_h$ is the usual vacuum-normalized path integral with a source $h(x)$
\eqn\path{\langle \CO_1(x_1)...\CO_n(x_n) \rangle_h={\int [{\cal D} \mu]  \CO_1(x_1)...\CO_n(x_n) e^{-S_0-\int d^dx h(x) \CO_0(x)}\over \int [{\cal D} \mu] e^{-S_0-\int d^dx h(x) \CO_0(x)}}~.}
The operators $\CO_i$ may or may not coincide with $\CO_0$.

Of particular interest are the correlation functions that can be obtained by derivatives of the disordered free energy, $W_D[g^i]$. In the context of second order phase transitions these correlation functions determine many aspects of the thermodynamics, and thus play an important role. Unlike the usual situation in CFTs where all correlation functions can be obtained from derivatives of the free energy, this is not the case for disordered theories.  We will be careful to illuminate these differences when they are important. For example, if we take the second derivative of the disordered free energy with respect to some couplings $g^i$ that couple to $\CO_i$ we would find the disorder-averaged connected correlation function 
\eqn\pathn{{\delta^2W_D[g^i]\over \delta g^i(x_1)\delta g^j(x_2)}= \overline{\langle\CO_i(x_1)\CO_j(x_2)\rangle}-\overline{\langle\CO_i(x_1)\rangle\langle\CO_j(x_2)\rangle }~.}
Note that the second term is not the same as $\overline{\langle \CO_i(x_1) \rangle} \cdot \overline{\langle\CO_j(x_2) \rangle}$. Generically the correlation functions \pathn\ behave differently from $ \overline{\langle\CO_i(x_1)\CO_j(x_2)\rangle}$. In the context of the holographic duality of field theories to gravitational theories, connected correlation functions of the type~\pathn\ naturally appear as derivatives of the bulk action of the gravitational theory.

The Harris argument~\Harris\ provides a simple diagnostic of when disorder is irrelevant. Suppose the scaling dimension of the scalar operator $\CO_0$ to which we couple disorder is~$\Delta$. Because of the quadratic term in $h$ on the right-hand side of~\disorder, it is natural to associate to $h$ dimension $d/2$, and to take $c$ to be a dimensionless coupling constant. Then, the term $\int d^dx h(x)\CO(x)$ in the action is relevant if $\Delta<d/2$, marginal if $\Delta=d/2$, and irrelevant if $\Delta>d/2$. Thus, we expect disorder to be innocuous for $\Delta>d/2$. This is the Harris bound. 
If $\Delta\leq d/2$ we have to analyze the situation case by case. In this situation, even arbitrarily small disorder may have profound effects on the low-energy behavior of the theory. It is also useful to keep in mind that the argument that for $\Delta > d/2$ disorder has no effect on the phase transition is only valid for small disorder; large disorder is similar to an irrelevant operator with a large coefficient, and it may have non-trivial consequences (another subtlety in this argument is considered in subsection~2.4).

One difficulty in analyzing the disordered free energy~\disorder\ by standard field theory methods is that we are averaging the logarithm of the partition function, and not the partition function itself, so we are not just adding extra terms to the action. One common way to address this is by appealing to the replica trick, where we average the $n$'th power of the partition function, and then perform an analytic continuation in $n$
\eqn\replica{\lim_{n\rightarrow 0}{Z^n-1\over n}=\log (Z)=W~.}
Alternatively, in some situations (including all the cases we analyze in this paper), the free energy can be sufficiently controlled directly, and then the replica trick is not necessary.

Analyzing disorder in free field theories (with disorder for the free scalar field) is relatively straightforward, and we begin by reviewing what is known about this case. We add some new observations on the nature of the low-energy fixed points (when these fixed points exist). In particular we show that they are not scale-invariant.

However, in many experimental situations one is interested in strongly coupled field theories, and then there are very few methods for analyzing disorder systematically. 
For example, consider the Ising model in $d$ dimensions with the exchange energy (the bond) having a small stochastic component of the type (1.2). This is described in the continuum limit by making the coefficient of the energy operator, $\epsilon(x)$, a random variable as above. For 
$d = 3$ this modifies the properties of the phase transition, and one finds new critical exponents. It is of great interest to find a useful approach to computing these critical exponents. 
One method that has been attempted is the epsilon expansion around $d=4$. It turns out that the expansion is in $\sqrt \epsilon$ and it unfortunately does not work as well as for the pure Ising model. See~\Folk\ for a relatively recent review and references. 

Another method that is useful when the disorder is close to being marginal is an expansion in $(d/2-\Delta)$ (for the Ising model $\Delta\sim 1.41$, and hence $d/2-\Delta\sim 0.09$ is numerically rather small and one can hope that expanding in it would give reasonably good results) -- one can view this as an expansion in the heat-capacity critical exponent. 
When $(d/2-\Delta)$ is small the disordered fixed point is sometimes close to the original, pure, fixed point. This method was used in two dimensions, see for example~\refs{\Ludwig,\DotsenkoIM,\DotsenkoSY,\ShimadaDM}, and an analysis in $2+\epsilon$ dimensions was considered in~\Cardy. The application of this approach to the random 3d Ising model will appear in a future publication (see~\Koma\ for some preliminary details). 

In this paper we study a different method for controlling the effects of disorder in interacting theories, which is to take the large-$N$ limit of vector models or of gauge theories.
In most of this paper we analyze large-$N$ theories where the coupling constant of a `single-trace operator' is taken to be a random variable. We analyze many aspects directly in field theory. In addition, if this theory 
has a known dual given by a weakly coupled gravitational theory on AdS$_{d+1}$, then 
$W[h, g^1,g^2,...]$ of~\disorder\ is equal to the AdS$_{d+1}$ bulk action with specified boundary conditions, and one can consider its average. If the gravitational theory is weakly curved then it is possible to explicitly study these disordered theories in AdS$_{d+1}$. This is done in section~5  (see~\refs{\Taka,\Niarcho} for earlier works, and see also~\ShangZW).\foot{See~\refs{\HartnollCUA,\HartnollFAA,\HartnollRZA} for the holographic analysis of disordered theories of the type that is relevant for quantum phase transitions, where disorder breaks the Poincar\'e symmetry. See also, for instance, \refs{\AreanOAA,\OKeeffeAWA,\BanksWHA} and references therein for additional recent studies of holographic disorder.} 

In the extreme large-$N$ limit, the field theory behaves as a `generalized free field theory', and we can exactly solve the path integrals that appeared above, and compute the correlation functions of the disordered theory.
We find again that scale-invariance of disordered fixed points is not guaranteed (even though they are fixed points of the renormalization group). We argue that when the disorder is marginal, it can sometimes be an exactly marginal deformation of the disordered fixed point, and that in some other cases the low-energy theory is scale-invariant but has unconventional dependence on the disorder. It would be interesting to understand if these non-standard properties of large-$N$ disordered fixed points can occur also for finite-$N$ theories. We also analyze what happens when a double-trace deformation is present in addition to the disorder.

The extreme large-$N$ limit is dual by the AdS/CFT correspondence to a free field theory on AdS$_{d+1}$, and in section 5 we use holography to make detailed computations and compare them to the field theory expectations. In particular, we reproduce the beta functions that the field theory predicts.

In sections 5 and 6 we study the leading $1/N$ corrections to the infinite $N$ results. When the disorder is of order one in the large-$N$ limit, these give small corrections that become important only at distances that (for marginal disorder) are exponentially large in $N$. However, these effects are important when 
 the disorder $c^2$ scales as the number of degrees of freedom (namely, as $N^2$ for gauge theories and as $N$ for
vector models). We cannot compute the $1/N$ corrections exactly, but we can analyze them perturbatively in the disorder, both holographically (where this involves an interacting field theory on AdS$_{d+1}$) and in field theory, and we do this at leading order. In general the behavior found for infinite $N$ is modified. At leading order in $1/N$ we find unconventional momentum-dependence in correlation functions of the disordered theory, with extra logarithms. Holography may allow for an exact analysis of this limit in some cases, but we leave this to future work.

\newsec{Disorder in Free Field Theories and in Large-$N$ Theories}

\subsec{A Review of Disorder in Free Field Theories}

We begin by analyzing disorder in free scalar field theories. This can be a first approximation to disorder in weakly coupled theories. It is also relevant for second order phase transitions when we have, on one side of the transition, a broken continuous global symmetry, since this gives a free theory of Nambu-Goldstone bosons at low energies.

Our action is 
\eqn\ff{S=\int d^dx\left(\half (\del_{\mu} \sigma(x))^2+h(x)\sigma(x)\right)~,}
and we take the coupling $h$ to vary randomly as in \prob. For specific $h(x)$ there is a competition between the fact that the
field $\sigma(x)$ wants to align itself with the local source $h(x)$, and the kinetic term that suppresses variations of the field.

For a given $h(x)$, we can compute the (normalized) two-point function of $\sigma(x)$ in closed form,\foot{The propagator in position space of a free field normalized as in \ff\  is 
\eqn\twopointf{\langle\sigma(x)\sigma(y)\rangle={\Gamma(d/2-1)\over 4\pi^{d/2}}{1\over (x-y)^{d-2}}~. } 
We omit these numerical prefactors and some combinatorial coefficients in this section in order not to clutter the equations.}
\eqn\twopointfree{\langle\sigma(x)\sigma(y)\rangle_h={1\over (x-y)^{d-2}}+\int d^dzd^dw {h(z)h(w)\over (x-z)^{d-2}(y-w)^{d-2}}~.}
 Thus, we find the exact result 
\eqn\havg{\overline{\langle\sigma(x)\sigma(y)\rangle}={1\over (x-y)^{d-2}}+\int d^dz{c^2\over (x-z)^{d-2}(y-z)^{d-2}}~.}
 The second term on the right-hand side is infrared divergent for $d\leq 4$. Therefore, the theory~\ff\ with $h(x)$ a random field is sick for $d\leq 4$. 
 This is the simplest example of a conformal field theory that ceases to exist because one of its couplings  is taken to be a random field \IM. In particular,
this computation shows that phase transitions with spontaneous breaking of continuous symmetries cease to exist in $d\leq 4$ once we make the couplings to the Nambu-Goldstone bosons  random~\IM. 

We can see another manifestation of the sickness encountered in~\havg\ by computing 
\eqn\onepointfree{\langle\sigma(x)\rangle_h=\int d^dz {h(z)\over (x-z)^{d-2}}~.} 
Therefore, 
\eqn\onepointavg{\overline{\langle\sigma(x)\rangle\langle\sigma(y)\rangle}=\int d^dz{c^2\over (x-z)^{d-2}(y-z)^{d-2}}~.} This is again infrared divergent for $d\leq 4$. 

 Note that the connected average 
$\overline{\langle\sigma(x)\sigma(y)\rangle}-\overline{\langle\sigma(x)\rangle \langle\sigma(y)\rangle}$ is altogether independent of disorder and finite! This is the combination that arises from the second derivative of the disordered free energy with respect to an extra source for $\sigma(x)$. 

However, it is possible to see the infrared catastrophe also at the level of the disordered free energy. To that end, we introduce a source for the composite operator $\sigma^2(x)$,
\eqn\ffi{S=\int d^dx\left(\half (\del\sigma(x))^2+h(x)\sigma(x)+g(x)\sigma^2(x)\right)~.}
When the CFT describes a second order phase transition, $g$ is proportional to $(T-T_c)$.
Averaging over the random field $h$ and computing the first derivative with respect to $g$ of the disordered free energy $W_D[g]$ we get 
\eqn\onepointss{-{\delta W_D\over \delta g(x)}=\overline{\langle\sigma^2(x)\rangle}=\overline{\int d^dyd^dy'{h(y)h(y')\over {(x-y)^{d-2} (x-y')^{d-2}}}   }=\int d^dy {c^2\over (x-y)^{2d-4}}~.}
 The integral is infrared divergent for $d\leq 4$. Hence, the operator $\sigma^2(x)$ does not exist as a local operator in the disordered theory, and the theory is thus sick. This is very similar to the familiar infrared divergence that leads to the Mermin-Wagner-Coleman theorem in $d=2$~\refs{\MW,\ColemanCI}. 

\subsec{The Low-energy Limit for $d>4$}

In the cases where the theories described in the previous subsection exist, namely $d>4$, it is interesting to ask what is their behavior at low energies.

Consider ${\delta^2 W_D\over \delta g(x) \delta g(y)}$, the connected averaged two-point function \eqn\connavg{\overline{\langle\sigma^2(x)\sigma^2(y)\rangle}-\overline{\langle\sigma^2(x)\rangle\langle\sigma^2(y)\rangle}~.} 
According to what we found above, this should be well defined for $d>4$. In second-order phase transitions we can interpret this correlation function as measuring the second derivative of the disordered free energy with respect to the temperature, namely the heat-capacity exponent.  One finds that in the deep infrared (compared to the scale set by $c$), and for $d>4$,
\eqn\heatcafree{\overline{\langle\sigma^2(x)\sigma^2(y)\rangle}-\overline{\langle\sigma^2(x)\rangle\langle\sigma^2(y)\rangle} \simeq {1\over (x-y)^{d-2}}\int d^d z{c^2\over (x-z)^{d-2}(y-z)^{d-2} } \simeq {c^2\over (x-y)^{2d-6} }~.}
The scaling dimension of $\sigma^2$ in the infrared is thus shifted (in the second derivative of the free energy) from $(d-2)$ to
$(d-3)$. As a result, the heat-capacity exponent of the disordered theory vanishes in $d=6$. In $d=6$, the two-point function~\heatcafree\ can be written as $\sim c^2 \int d^6p e^{ipx} \log(p^2) $. This ``dimensionless'' behavior in $d=6$ is the starting point of a systematic $\epsilon$-expansion of the disordered free energy of interacting theories  around $d=6$~\AYS.
It is important to point out that again the connected average~\heatcafree\ behaves very differently from $\overline{\langle\sigma^2(x)\sigma^2(y)\rangle}$, which is dominated, like the disconnected contribution $\overline{\langle\sigma^2(x)\rangle\langle\sigma^2(y)\rangle}$, by a term that scales as $c^4/(x-y)^{2d-8}$ at large distances. 

Let us also discuss the connected averaged three-point function of $\sigma^2$, namely, ${\delta^3 W_D\over \delta g(x) \delta g(y)\delta g(z)}$. One finds that it receives a contribution that goes like $c^2$ from disorder, but all the contributions of order $c^4$ and higher exactly cancel:
\eqn\threept{\eqalign{&-{\delta^3 W_D\over \delta g(x) \delta g(y)\delta g(z)}\simeq {1\over (x-y)^{d-2}(x-z)^{d-2}(y-z)^{d-2}  }\cr&\qquad\qquad\qquad\qquad\qquad+\left[{c^2\over (x-y)^{d-4}(x-z)^{d-2}(y-z)^{d-2}  }+{2 \ {\rm permutations}}\right]~.}}
In the infrared the terms proportional to $c^2$ always dominate. As expected, for $d>4$ the correlation function~\threept\ decays at long distances and so the theory is well behaved. However, in the deep infrared the three-point function~\threept\ is inconsistent with conformal invariance. In fact, it is not even consistent with scale-invariance.\foot{Note that in momentum space there is a well-known relation \refs{\IM,\Grinstein,\AYS,\Young,\ParisiKA} between
the IR limit of disorder-averaged connected correlation functions of the $d$-dimensional
free theory, such as \heatcafree\ and \threept, and connected correlation functions
of the pure $(d-2)$-dimensional free theory. This relation holds when all
momenta lie in a $(d-2)$-dimensional subspace, and it continues to hold
to all orders in perturbation theory (though it does not hold non-perturbatively).
This relation implies that these specific disorder-averaged connected
correlation functions in momentum space obey a $(d-2)$-dimensional
scaling symmetry (and even a conformal symmetry), as do the
correlation functions \heatcafree\ and \threept\ when integrated over two of the
dimensions. It also implies some relations between the critical exponents of
the $d$-dimensional disordered theory and the $(d-2)$-dimensional pure
theory. However, the full correlation functions of the disordered free theory do not obey any
scaling symmetry, as exhibited already by the form of \heatcafree\ and \threept.} Based on the two-point function~\heatcafree\ one would have expected that $\sigma^2$ carries dimension $d-3$, but the overall dimension of the three-point function~\threept\ in the infrared is $3d-8$ rather than $3d-9$
\foot{Another option is to force the connected correlation functions to be
scale-invariant in the infrared, by scaling each operator according to
the scaling dimension implied by its two-point function. With this scaling
we find that all connected $n$-point functions with $n>2$ (such as \threept)
vanish in the IR. This would imply that the disorder-averaged IR theory is non-local.
For instance, all connected 3-point functions $\overline{\vev{\sigma(x) \sigma(y)
{\cal O}(z)}}$ with any operator ${\cal O}$ would vanish in the IR, implying that $\sigma(x)
\sigma(y)$ has no overlap with any local operator (except the
identity operator); this is unlike the situation in the UV theory
where the connected $\vev{\sigma(x) \sigma(y) \sigma^2(z)}$ is non-zero. The same
is true also for any other product of two operators ${\cal O}_1(x) {\cal O}_2(y)$
in the IR. Thus, the interpretation of having no scale-invariance in the IR seems more
reasonable to us. We thank S. Rychkov for discussions on this
point.}.

In the next section we will see how these results are reproduced using the replica trick, and we will return there to the discussion of scale-invariance in the infrared. Here we only mention that this non-scale-invariance in the infrared can be traced in the replica trick computation to the fact that the replicated theory is unstable; the deep infrared limit and the $n\rightarrow 0$ limit of~\replica\ do not commute.

\subsec{Disorder in the large-$N$ limit}

Large-$N$ field theories (such as $SU(N)$ gauge theories in the 't Hooft limit or $O(N)$ vector models) provide useful examples where disorder can be analyzed explicitly in the $N\to \infty$ limit, since connected correlation
functions in these theories are suppressed by powers of $1/N$. 

The simplest case arises when we have disordered couplings for
'single-trace operators' (operators invariant under $SU(N)$ or $O(N)$ that cannot be written as products of invariant operators) in a large-$N$ theory and we normalize their two-point functions and sources to be
of order one in the large-$N$ limit; for infinite $N$ this leads to `generalized free fields'.
Generalized free fields are CFT operators that obey the same statistics as free fields, even though they can have different dimensions from free fields. We normalize the two-point function of a scalar operator $\CO(x)$ of dimension $\Delta$ to 
\eqn\singletwo{\langle \CO(x)\CO(y)\rangle={1\over (x-y)^{2\Delta}}~,}
and higher-point functions are then given by Wick's theorem. The only operators in such theories are products of $\CO$
and of derivatives of $\CO$.
As a consequence, the OPE of $\CO(x)\CO(y)$ does not include the energy-momentum tensor (unless $\Delta=(d-2)/2$, in which case $\CO$ is an ordinary free field). Thus, one cannot think of such a theory as a local CFT, but one can think of it as a decoupled sector of a local CFT
with $c_T=\infty$, where $c_T$ is the central charge appearing  in the two-point function of the energy-momentum tensor.
In the AdS/CFT correspondence, a generalized free field is related to a free field of mass $m$ in an $AdS_{d+1}$ space of radius $L$, with $\Delta(\Delta-d) = m^2 L^2$.\foot{Such relations between sectors of CFTs and field theories on AdS space can exist also for non-free field theories on $AdS_{d+1}$ \AharonyZEA.} 

In this section we study such theories with arbitrary values of $\Delta < d/2$, and with disorder for $\CO$. In this case we can compute everything exactly as a function of $c^2$, as long as $c^2$ remains finite in the large-$N$ limit. As in the previous subsection, we find a non-trivial low-energy limit with novel behavior. In the next section we will discuss the
specific case of $\Delta=d/2$, where some new features arise. In large-$N$ disordered theories the free energy and correlation functions can be exactly computed as a function of the interesting couplings, so we do not need to use the replica trick. However, in the next section we will perform computations also using the replica trick, and we will see that in some cases there are subtleties in using the replica trick for generalized free fields.

The computations in this subsection are straightforward generalizations of the computations performed above for free fields.
Consider a generalized free field $\CO$ with $\Delta < d/2$, so that disorder is a relevant perturbation. Its two-point function in the presence of some source $h$,
\eqn\gfftwo{\langle\CO(x)\CO(y)\rangle_h={\int [D\mu]\CO(x)\CO(y)e^{-S_0-\int d^dx h\CO}   \over \int [D\mu]e^{-S_0-\int d^dx h\CO}},}
can be solved exactly. In momentum space we get (up to unimportant prefactors that we will ignore throughout this section)
\eqn\twptge{\langle\CO(k)\CO(l)\rangle_h=k^{2\Delta-d}\delta^{(d)}(k+l)+k^{2\Delta-d}l^{2\Delta-d}h(k)h(l)~,}
and in position space
\eqn\twptge{\langle\CO(x)\CO(0)\rangle_h={1\over x^{2\Delta}}+\int d^dzd^dw{h(w)h(z)\over(x-z)^{2\Delta}w^{2\Delta}}~.}
We can now directly average these expressions using~\prob\ to find an exact result \Taka
\eqn\finalii{\bar{\langle\CO(k)\CO(-k)\rangle}={(k^2)^{d/2-\Delta}+c^2\over (k^2)^{d-2\Delta}    }   ~. }
 
In the deep UV we find 
\eqn\uvgffvev{\lim_{k^2\rightarrow \infty}  \bar{\langle\CO(k)\CO(-k)\rangle}  = (k^2)^{\Delta-d/2}\longrightarrow 
\lim_{x\rightarrow 0}  \bar{\langle\CO(0)\CO(x)\rangle} = x^{-2\Delta},}
and in the deep infrared
\eqn\irgffvev{\lim_{k^2\rightarrow 0}  \bar{\langle\CO(k)\CO(-k)\rangle}  = { c^2\over  (k^2)^{d-2\Delta}}\longrightarrow 
\lim_{x\rightarrow \infty}  \bar{\langle\CO(0)\CO(x)\rangle} =c^2x^{d-4\Delta}.}
We see that in the UV, there is no effect of the weak disorder, and the two-point function in the UV is the same as in the pure theory, as expected
when the disorder is relevant. In the infrared, however, the disorder term dominates and the behavior is altered. 

The distribution $1/(k^2)^{d-2\Delta}$ is non-integrable for $d-4\Delta\geq 0$, i.e $\Delta\leq d/4$. Exactly when this happens the correlation function in position space does not decay. This signifies a disease of the theory which generalizes the result of~\IM\ (our computation reduces to~\IM\ when $\Delta=(d-2)/2$, as reviewed in subsection 2.1).
Note that if the original CFT is unitary, then $\Delta\geq d/2-1$, and thus $d-4\Delta\geq 0$ can only happen if $d\leq 4$. 

Let us next consider the correlation function \finalii\ from which we subtract the piece which is disconnected before the $h$-average is taken. Indeed, $\langle\CO(k)\rangle_h=k^{2\Delta-d}h(k)$, so $\overline{\langle\CO(k)\rangle \langle\CO(l)\rangle}=c^2k^{4\Delta-2d}$. 
Hence, the piece that is dominant in the infrared in \finalii\ exactly cancels with the average of the disconnected correlation function, and
the second derivative of the free energy behaves as
\eqn\connected{\bar{\langle\CO(k)\CO(-k)\rangle}-\bar{\langle\CO(k)\rangle\langle\CO(-k)\rangle}={(k^2)^{\Delta-d/2}}~, }
which is completely independent of disorder! It is easy to see that the same is true also for higher derivatives of the free energy, and even for connected correlators of $\CO$ with other operators. This is due to the fact that all the connected correlators of more than  two generalized free fields vanish.

As in subsection~2.1, to see the sickness of the theory for $\Delta \leq d/4$, we can also add a source for the operator $\CO^2$.  Then we can take a derivative of the free energy with respect to the source and obtain 
\eqn\otwovev{
\vev{\CO^2(k)}_h=\int d^dl{h(k-l)h(l)\over (k-l)^{d-2\Delta}l^{d-2\Delta}}.   } 
If we average over $h$, only the zero momentum piece has a non-zero average (as expected from translation invariance of the statistical ensemble) and we find 
\eqn\otwoavvev{
\overline{\vev{\CO^2(k=0)}}=c^2 Vol(\IR^d)\times \int d^dl {1\over l^{2d-4\Delta}}.}
This is infrared divergent if $\Delta\leq d/4$. So we could say that the operator $\CO^2$ does not exist in the disordered theory if $\Delta\leq d/4$. For example, consider the free (non-critical) $O(N)$ vector model at very large $N$. The operator $\CO=\vec\phi^2$ is a dimension $\Delta=d-2$ generalized free field operator. Its dimension is smaller than $d/4$ for $d\leq 8/3$. Hence, in this range, introducing small disorder in $\vec\phi^2$ would render the theory ill-defined.

 Now we will assume $d/2>\Delta>d/4$ and compute the infrared anomalous dimension of $\CO^2$. We will study the averaged connected two-point function. This subtracts the most infrared divergent pieces and has a clear physical interpretation as the second derivative of the free energy with respect to some source. 
The final answer for the average of the second derivative of the free energy with respect to the source of $\CO^2$ is 
\eqn\dtvi{\bar{\langle\CO^2(k)\CO^2(-k)\rangle}- \overline{\langle\CO^2(k)\rangle\langle\CO^2(-k)\rangle}  =(k^2)^{-d/2+2\Delta}+c^2(k^2)^{-d+3\Delta}~.}
There are no higher corrections in $c^2$. The dominant term in the UV is the first one which leads to the dimension $2\Delta$, as appropriate in a generalized free field theory. 
The infrared dominant term is the second term. Thus, $\CO^2$ behaves as a field with a shifted dimension: 
\eqn\scalingtran{\Delta_{\CO^2}: \qquad\qquad 2\Delta~\longrightarrow ~3\Delta-d/2~.} 
As expected, the marginal case $\Delta=d/2$ is a fixed point of the transformation. Also, the dimension of $\CO^2$ always decreases under such  renormalization group (RG) flows.

However, the infrared theory does not adhere to the rules of any standard scale-invariant fixed point. Indeed, if we consider higher derivatives of the disordered free energy with respect to the source of $\CO^2$, they are all at most linear in $c^2$, and we get in the deep infrared 
higher correlation functions of $\CO^2$ which do not respect the scaling~\scalingtran. For example, generalizing \threept, the averaged connected three-point function scales in the infrared as \eqn\disthree{\overline{\langle \CO^2(x)\CO^2(y)\CO^2(z) \rangle^{conn.}}\sim {c^2\over (x-y)^{4\Delta-d}(y-z)^{2\Delta}(x-z)^{2\Delta}}+{\rm 2\ permutations}~.}
Such a three-point function is certainly inconsistent with $\CO^2$ being a primary of the conformal group, and, given~\scalingtran, it is even inconsistent with $\CO^2$ being an eigen-operator of the dilatation operator in a scale-invariant field theory. As in the free field theory discussed above, one can
also interpret the theory as having scale-invariance but
with all higher-point functions vanishing, but this seems to lead
to a loss of locality. We will 
discuss the interpretation of these results in the replica trick computation in the next section. 

\subsec{Generalized Free Fields with $\Delta>d/2$}

In this case disorder is irrelevant according to the Harris criterion so it would appear that there is nothing to be discussed. We will see below that this is not quite the case. The momentum space two-point correlation function in the pure theory is given by 
\eqn\puretwo{\langle \CO(k)\CO(-k)\rangle=k^{2\Delta-d}~.}
However, one can also add a constant which corresponds to a contact term in position space 
\eqn\modifiedprop{\langle \CO(k)\CO(-k)\rangle=\Lambda^{2\Delta-d}_{UV}+k^{2\Delta-d}~,}
where $\Lambda_{UV}$ is some unknown UV scale which is scheme-dependent (for instance, it can depend on the details of an underlying lattice theory).\foot{If one imagines coupling $\CO(x)$ to a source $h(x)$, the above contact term is obtained by adding to the Lagrangian $\sim \Lambda_{UV}^{2\Delta-d}\int d^dx h^2(x)$. Such a contact term would be generically induced by an RG flow. We will see in section 4 that this indeed happens in the double-trace flow.}

If we deform our theory by disorder for $\CO$, then, for instance, the disordered two-point function \finalii\ now becomes (up to unimportant numerical coefficients)
\eqn\twopointnoHarris{\overline{\langle  \CO(k)\CO(-k)\rangle} = \Lambda^{2\Delta-d}_{UV}+k^{2\Delta-d}+c^2 \left(\Lambda^{2\Delta-d}_{UV}+k^{2\Delta-d}\right)^2~.}
Keeping the factors that are most important in the infrared we obtain a contact term plus $(1+2c^2\Lambda^{2\Delta-d}_{UV})k^{2\Delta-d}+\cdots$. As long as $\Lambda_{UV}\neq0$ the effects of disorder are thus not suppressed compared to the pure theory.\foot{In position space, the contribution containing $c^2\Lambda_{UV}^{2\Delta-d}$ arises from the integrated disorder operator coinciding with one of the insertions in the correlation function.} On the other hand, if we choose $\Lambda_{UV}=0$, the term proportional to $c^2$ would be multiplied by $k^{4\Delta-2d}$, which is parametrically suppressed in the infrared. One can verify that many other correlation functions (including connected averaged correlation functions) are affected by disorder, and that the effects of disorder cannot be removed by a redefinition of the operators when $\Lambda_{UV}\neq 0$.

We thus see that although the disorder is irrelevant in the technical sense, it does modify correlation functions in the infrared. (This is perhaps reminiscent of dangerously irrelevant operators in standard renormalization group flows. The difference is that here the effects of irrelevant disorder are only present when we allow for a contact term as in~\modifiedprop.)

\newsec{Disorder at Large-$N$ using the Replica Trick}

\subsec{Large-$N$ Theories with Disorder for an Operator with $\Delta=d/2$}

In the marginal case $\Delta=d/2$  it is natural to compute the beta function for the disorder. We can compute this using the replica trick~\replica. The approach we use below is analogous to that of~\refs{\Ludwig,\Cardy,\DotsenkoIM,\DotsenkoSY,\ShimadaDM}. The results are however not the same because in the context of large-$N$ theories double-trace operators play an important role.

We start by considering 
\eqn\replica{W_n\equiv \int [{\cal D}h] Z^n[h] e^{-\int d^dx {h(x)h(x)\over 2c^2}},}
where 
\eqn\zh{Z[h]=\int [{\cal D}\mu]e^{-S-\int d^dx h(x)\CO(x)}.}
We can then compute the averaged disordered free energy \disorder\ by 
\eqn\continuation{W_D={d\over  dn} W_n\biggr|_{n=0}~.}
We write 
\eqn\npower{Z^n[h]=\int \prod_{A=1}^n [{\cal D}\mu_A]e^{-\sum_AS_A-\sum_A\int d^dx h(x)\CO_A(x)}~,  }
where capital Latin indices go over the replicas, $A=1,\cdots,n$. We therefore have the path integral 
\eqn\npowerpi{W_n=\int [{\cal D}h] e^{-\int d^dx {h(x)h(x)\over 2c^2}} \prod_{A=1}^n [{\cal D}\mu_A]e^{-\sum_AS_A-\sum_A\int d^dx h(x)\CO_A(x)}~.}
We solve the $h$ path integral first. The equation of motion of $h$ sets $h(x)=-c^2\sum_A\CO_A(x)$ and thus
\eqn\npowerpi{W_n=\int\prod_{A=1}^n [{\cal D}\mu_A]e^{-\sum_AS_A+{c^2\over 2}\int d^dx \sum_{A,B}\CO_A(x)\CO_B(x)}~.}
Therefore, we have a collection of $n$ identical CFTs perturbed by some marginal couplings that connect them. 

In \npowerpi\ one has to decide what to do with the operator $\CO_A(x)\CO_A(x)$ for some particular $A$. In generic field theories
there is no marginal operator appearing in this Operator Product Expansion (OPE), so this term can just be dropped.
However, in large-$N$ theories (generalized free field theories) the OPE takes the form
\eqn\OPEgenfree{\CO(x)\CO(0)\sim {1\over x^d}+{\CO^2(0)}+\cdots~.}
Thus, we simply interpret the product $\CO_A(x) \CO_A(x)$ as the operator $\CO^2_A(x)$. With this interpretation we have
(keeping only marginal terms)
\eqn\npowerpii{W_n=\int\prod_{A=1}^n [{\cal D}\mu_A]e^{-\sum_AS_A+{c^2\over 2}\int d^dx \sum_{A,B}\CO_A(x)\CO_B(x)}~.}
Expanding~\npowerpii\ in weak disorder, we get 
\eqn\expansion{\eqalign{&W_n= \int\prod_{A=1}^n [{\cal D}\mu_A]e^{-\sum_AS_A} \cr&\left[1+{c^2\over 2}\int d^dx \sum_{A, B}\CO_A(x)\CO_B(x)+{c^4\over 8}\int d^dxd^dy \sum_{A,B,C,
D}\CO_A(x)\CO_B(x)\CO_C(y)\CO_D(y)+\cdots\right] ~. }} 
The third term in the second line can renormalize the second term in the second line. Indeed, marginal terms in the OPE $\left(\CO_A\CO_B\right)(x)\left(\CO_C\CO_D\right)(y)$ lead to logarithmic divergences at short distances, which can be absorbed either by renormalizing $c^2$ or by adding new terms to the action.

In large-$N$ theories there is just one way to get a marginal term as $x\to y$ in the third term of \expansion, i.e. from the unit operator in a single contraction of $\CO$'s. This gives
\eqn\OPEdt{(\CO_A \CO_B)(x)(\CO_C \CO_D)(0)\sim {1\over x^d} \left(\delta_{AC}\CO_B\CO_D+{\rm 3\ permutations}\right),}
and
\eqn\OPEdtt{\sum_{A,B}\CO_A \CO_B(x)\sum_{C,D}\CO_C \CO_D(0)\sim {4n\over x^d}\sum_{A,B}\CO_A \CO_B(0)~.}
The integral $d^d x$ over \OPEdtt\ in \expansion\ diverges logarithmically, so the conclusion is that we have a beta function\foot{Similar methods were applied in $d=2$ to describe the disordered Ising and Potts models, see. e.g.~\refs{\Ludwig,\DotsenkoIM,\DotsenkoSY,\Cardy,\ShimadaDM}. }
\eqn\betafun{{dc^2\over d\log(\mu)}=-\gamma(d) c^4 n +O(c^6)~,}
where $\gamma(d) \equiv {\rm Vol}(S^{d-1}) = 2 \pi^{d/2} / \Gamma(d/2)$.

The beta function for the physical disordered theory is obtained by simply substituting $n=0$ in~\betafun, letting the derivative in \continuation\ act on the first line of \expansion, so
\eqn\betagff{{dc^2\over d\log(\mu)}=0~.}
This shows that in generalized free field theories the marginal disorder $c$ is exactly marginal!

At first sight this is not too surprising, since we found in the previous section that the averaged connected two-point function of $\CO$ does not depend on the disorder (so one could say that $\CO$ does not get renormalized as a result of disorder). However, the averaged connected correlation function of double-trace operators now takes the form
\eqn\gffdoubletwo{\overline{\langle \CO^2(x)\CO^2(0)\rangle}-\overline{\langle \CO^2(x)\rangle\langle \CO^2(0)\rangle}={2\over x^{2d}}+{4\over x^{d}}c^2\int d^dy{1\over y^{d}(y-x)^{d}}~.}
Equivalently, in momentum space,  
\eqn\momdouble{\eqalign{&\overline{\langle \CO^2(k)\CO^2(-k)\rangle}-\overline{\langle \CO^2(k)\rangle\langle \CO^2(-k)\rangle}=\cr& \qquad 2\int d^dq \log(q^2/\mu^2)\log((k-q)^2/\mu^2)+4c^2 \int d^dq \log(q^2/\mu^2)\log^2((k-q)^2/\mu^2)~.}}
Clearly the first term makes sense because if we rescale $\mu$ we get an integral that does not depend on $k$ and hence a contact term. But the second integral transforms under rescaling $\mu$ non-trivially already at separated points. Therefore, the two-point correlation function looks reminiscent of a logarithmic CFT (logarithmic CFTs are argued in~\CardyRQG\ to arise naturally in disordered theories):
\eqn\momdoublei{\eqalign{&\overline{\langle \CO^2(x)\CO^2(0)\rangle}-\overline{\langle \CO^2(x)\rangle\langle \CO^2(0)\rangle}={2+4c^2\log(x^2\mu^2)\over x^{2d}}~.}}
$\CO^2$ may seem to have a non-zero anomalous dimension as a function of the exactly marginal disorder $c^2$, but the result \momdoublei\ is actually exact for any $c^2$ so it does not exponentiate as an anomalous dimension should. 

To summarize, the exactly marginal parameter $c^2$ does have an effect on the correlation functions.

\subsec{Exact Replica Computations and an Interpretation of non-Scale-Invariance}

Next, we want to understand from the point of view of the replica trick
why the disorder limit $n\rightarrow 0$ of \npowerpii\ leads to a non-scale-invariant theory for the case of $\Delta < d/2$. We can solve the replicated generalized free field theory \npowerpii\ exactly for any $n$. For any integer $n\neq 0$ we have a standard QFT so we expect the low-energy limit to be scale-invariant, but this is not what we found for the disordered theory in the previous section. We will see that the limit of $n\rightarrow 0$ does not commute with the long distance limit in the class of theories we study here.\foot{Even for integer $n$ the replicated theory is not really a standard QFT in the case of generalized free fields, because of an instability discussed below.} 

We study the theory of $n$ initially decoupled generalized free field theories with an interaction $(-\half c^2 \sum_{A,B} \CO_A \CO_B)$, where $A,B=1,\cdots,n$. We will calculate the correlation function $\langle\sum_A \CO_A(x)\sum_B \CO_B(0)\rangle$ exactly. By the time we take $n\rightarrow 0$, the derivative of this with respect to $n$ gives the two-point connected average of the generalized free fields in the disordered theory. Note that this correlation function in the replicated theory is not normalized, so we always have to multiply the answers below by $\langle 1\rangle$ at the end of the day. That, however, will not change our conclusions. After some combinatorics we find 
\eqn\fullrep{\langle\sum_A \CO_A(k)\sum_B \CO_B(-k)\rangle={n\over k^{d-2\Delta} -nc^2    }~.}
This two-point function signals an instability because of the singularity in the denominator. This is not surprising as the interaction term that we must add in the replicated theory, $\half c^2 \sum_{AB} \CO_A \CO_B$, appears with the wrong sign in the potential \npowerpi. Now:

\medskip 

\item{A.} If we first apply to \fullrep\ the operator ${d\over dn}\bigr|_{n=0}$ we clearly get 
\eqn\ddnav{\overline{ \langle \CO(k) \CO(-k) \rangle^{conn.} }={1\over k^{d-2\Delta}}~.}
This corresponds to $\CO$ having scaling dimension $\Delta$ (in agreement with~\connected). Note that it is completely independent of disorder, as we found above.
\item{B.} If, on the other hand, we first take the limit of small momentum (crossing the singularity at $k^{d-2\Delta}=nc^2$), we get 
\eqn\repvev{\langle\sum_A \CO_A(k)\sum_B \CO_B(-k)\rangle=-{1\over c^2}-{1\over n c^4}k^{d-2\Delta}+\cdots~.}
The first term is a contact term, and the other terms have
singular dependence on $n$ for small $n$. This formula naively implies that the infrared dimension of $\CO$ is $d-\Delta$, which is what we are used to from double-trace deformations (though for any $n\neq 1$ the deformation is not equivalent to a double-trace deformation).

\medskip

We therefore clearly see that the limit of small $n$ and the infrared limit do not commute. This is due to the singularity in~\fullrep, which signals an instability at lower and lower energies as $n\to 0$. 
If we go to very small momenta first, we go below the unphysical pole (whose location scales as a positive power of $n$). But if we take $n\rightarrow 0$ first, the instability is ``sequestered,'' and we get the correct answer for the  disordered behavior.

The analysis is particularly simple for free scalar fields as in subsection 2.1. In this case the replicated theory is again a free field theory, with one
field of mass $m^2 = - n c^2$, and with $(n-1)$ massless free fields. The presence of the tachyonic field for any finite $n$ indicates that the theory
is unstable, and this becomes problematic once we go to momenta below $k^2 = n c^2$.

We now turn to the correlator $\langle \sum_A\CO_A^2(k) \sum_B\CO_B^2(-k)\rangle$. After some combinatorics one arrives at the expression 
\eqn\double{\eqalign{&\langle \sum_A\CO_A^2(k) \sum_B\CO_B^2(-k)\rangle=2\int d^dl\biggl(n l^{2\Delta-d} (k-l)^{2\Delta-d}+2nc^2 (k-l)^{2\Delta-d}{l^{4\Delta-2d}\over1-nc^2l^{2\Delta-d}  } \cr& \qquad \qquad \qquad \qquad \qquad\qquad\qquad\qquad +n^2 c^4 {l^{4\Delta-2d}\over 1-nc^2 l^{2\Delta-d} } {(k-l)^{4\Delta-2d}\over 1-nc^2 (k-l)^{2\Delta-d} }\biggr)~.}}
Now we demonstrate again that the infrared limit and the small $n$ limit do not commute:

\medskip

\item{A.} If we apply the operator ${d\over dn}\bigr|_{n=0}$ first we get 
\eqn\doublei{\eqalign{&\overline{\langle \CO^2(k) \CO^2(-k) \rangle}=2\int d^dl\biggl(l^{2\Delta-d} (k-l)^{2\Delta-d}+2c^2 (k-l)^{2\Delta-d}l^{4\Delta-2d} \biggr)~.}}
This coincides with~\dtvi, and the infrared dominant term is the one proportional to $c^2$, which gives $k^{6\Delta-2d}$, that translates in position space to $x^{d-6\Delta}$. This is precisely the transformation~\scalingtran.
\item{B.} If we take the infrared limit first, we notice that for instance for $n=1$ the leading infrared term cancels between the three terms in~\double. This is important, because in this case where we just have an almost standard double-trace deformation, we indeed expect that the dimension of $\sum_A\CO_A^2$ should be $2d-2\Delta$. For any other $n> 0$ the cancellation does not happen and we get that the leading infrared behavior is $k^{4\Delta-d}$, corresponding to $x^{-4\Delta}$ in position space, so that the dimension of $\sum_A \CO_A^2$ stays $2\Delta$. 

\medskip

It would be interesting to understand if the breakdown of scale-invariance in disordered theories is always associated to such instabilities in the replicated theory. 

Preliminary studies suggest that the issues of non-scale-invariance~\disthree\ and logarithmic correlation functions as in~\momdoublei\ can also be naturally understood by generalizing the Osborn equations~\OsbornGM\ to disordered systems. However, we leave this topic for the future.

\newsec{Double-Trace Deformations and Disorder}

We now generalize the study above to include a double-trace operator $\CO^2$ in the action in addition to the disorder. The reasons for doing that are twofold:
\item{1.} In a general field theory we expect that such double-trace deformations should be present in any physical application (since the double-trace deformation is relevant at large $N$ if and only if disorder is relevant). 
\item{2.} We have seen above that scale non-invariance is related to an instability of the replicated theory. We would like to test this by curing the instability with an explicit double-trace deformation. 

\medskip

Our theory is thus 
\eqn\lambdah{S=S_0+\half \lambda \int d^dx \CO^2(x)+\int d^dx h(x) \CO(x)~,}
with $h$ a random Gaussian field. Repeating the steps near~\finalii\ we find that the two-point function now takes the exact form \Taka
\eqn\finaliii{\eqalign{&\bar{\langle\CO(k)\CO(-k)\rangle}={1\over (k^2)^{d/2-\Delta}+\lambda}+{c^2\over \left((k^2)^{d/2-\Delta}+\lambda\right)^2   }~.   } }
For $\lambda=0$ it coincides with~\finalii,
and for $c=0$ it coincides with the familiar results in theories with double-trace deformations (see e.g.~\GubserVV).

We assume $\Delta < d/2$. At very high energy we always asymptote to $(k^2)^{\Delta-d/2}$, which corresponds to $\CO$ having dimension $\Delta$. This is because the disorder and double-trace deformations are both small in the ultraviolet.

Let us now compute the averaged connected two-point function. We subtract the average of one-point functions from~\finaliii\ to find
\eqn\modpro{\eqalign{&\overline{\langle\CO(k)\CO(-k)\rangle}-\overline{\langle\CO(k)\rangle\langle\CO(-k)\rangle}= {1\over k^{d-2\Delta}+\lambda}~. } }
 So we see that the connected averaged two-point function does not see the disorder at all, as in the case of $\lambda=0$~\connected. 

As before, we can also consider the two-point function of $\CO^2$, removing the pieces which are disconnected before the $h$-average. One can find exact results even in the presence of the double-trace coupling. Our final answer is 
\eqn\twopointdouble{\overline{\langle\CO^2(k)\CO^2(-k)\rangle^{conn.}}=\int d^dl\left[ {1\over l^{d-2\Delta}+\lambda}  {1\over (k-l)^{d-2\Delta}+\lambda} +c^2\left( {1\over l^{d-2\Delta}+\lambda}\right)^2  {1\over (k-l)^{d-2\Delta}+\lambda} \right]~.  }
This coincides with the previous result \dtvi\ for $\lambda=0$, where the dimension in the infrared is given by $3\Delta-d/2$. Another special case is $c=0$. In this case at large $k$, the virtual momentum $l$ is going to be large as well, and we can neglect $\lambda$ and get $\int d^dl {1\over l^{d-2\Delta}(k-l)^{d-2\Delta}}\sim{1\over k^{d-4\Delta}}\rightarrow {1\over x^{4\Delta}}$. This power-law behavior is associated to an operator of dimension $2\Delta$, as expected from $\CO^2$ at high energies. At low energies all the momenta are small and we expand and get 
$\int d^dl\left[{1\over \lambda^2}+{l^{d-2\Delta}\over \lambda^3}+{l^{d-2\Delta}(k-l)^{d-2\Delta}\over \lambda^4}\right] +\cdots$. The first two terms do not give nontrivial momentum-dependence in the infrared, and so the leading piece comes from the third term, which gives $k^{3d-4\Delta}\rightarrow x^{-4d+4\Delta}$.
Hence, the infrared dimension of the operator $\CO^2$ is given by $2d-2\Delta$, as we expect for the ordinary double-trace deformed theory. 

Analyzing~\twopointdouble\ for generic non-zero $c$ and $\lambda$, it turns out that the infrared scaling is always~$x^{4\Delta-4d}$, so that $\CO^2$ has dimension $2d-2\Delta$. This has a simple interpretation. If we imagine first reaching the fixed point of the double-trace deformed theory, then $\CO$ has dimension $d-\Delta$. Therefore, according to the Harris criterion, the disorder is now irrelevant, and hence the two-point function after turning on disorder has the same IR scaling as in the ordinary double-trace deformed theory. Therefore at the level of two-point functions the qualitative picture of the renormalization group flow is as follows:

\medskip

\medskip
\epsfxsize=3.4in \centerline{\epsfbox{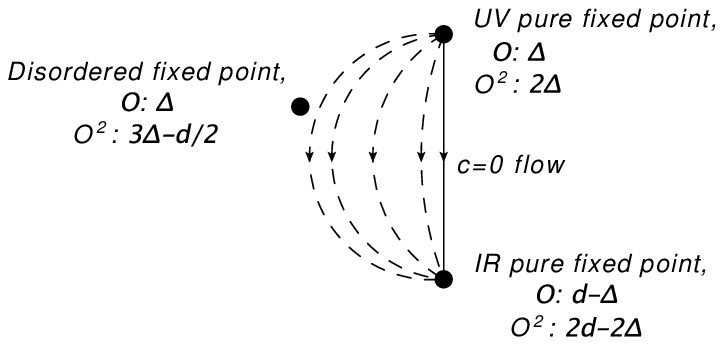}}
%\nobreak\vskip-0.8cm {
{\it \vbox{ {\bf Figure 1:}
{\it A schematic picture of the large-$N$ renormalization group flows in the presence of disorder and a double-trace deformation.}}}
\medskip

However, in view of subsection 2.4, this is not quite the whole story. In higher-point functions we now find results that are consistent with scale-invariance but depend non-trivially on the dimensionless parameter $c^2/\lambda$. At long distances we have at separated points an exact result
\eqn\npointdouble{\overline{\langle\CO^2(x_1) \cdots \CO^2(x_n)\rangle^{conn.}} = \langle\CO^2(x_1) \cdots \CO^2(x_n)\rangle^{conn.}_{c=0} \left(1 + 2{ c^2 \over \lambda} \right)~.}
The dependence on $c^2/\lambda$ cannot be removed by a redefinition of the operators. From the point of view of the infrared double-trace fixed point, $\lambda$ is a cutoff scale in the ultraviolet. Even though $c^2$ is technically irrelevant in the double-trace fixed point, the additional dependence on $c^2/\lambda$ in~\npointdouble\ is present because the RG flow to the double-trace fixed point induces a contact term as in \twopointnoHarris.

Let us note that in the presence of the double-trace deformation, the instability found in~\fullrep\ is removed. Given any $\lambda>0$, for small enough but finite $n$, there is no unphysical singularity, because the denominator in~\fullrep\ is modified as $cn^2\rightarrow cn^2-\lambda$. This is in accord with the fact that scale-invariance is not violated by disorder in the presence of $\lambda$.

\subsec{Marginal Double-Trace Deformations and Disorder}

If we assume that the generalized free field $\CO$ has $\Delta=d/2$, then disorder is marginal and also the double-trace deformation is marginal. We should thus study the beta function in this case, as we did in subsection~3.1. 

In the replica trick, we should now deform the $n$ decoupled theories by 
\eqn\dt{\delta S = {\lambda\over2} \sum_A\CO_A^2-{c^2\over 2}\sum_{A,B}\CO_A\CO_B~.}
The sign of the double-trace deformation is the usual one necessary for stability if $\lambda>0$. Simple combinatorics shows that 
the singular terms in the OPE take the form
\eqn\comps{\eqalign{&\sum_A\CO_A^2(0)\sum_B\CO_B^2(x) \simeq {4\over x^d} \sum_A \CO_A^2(0)~,\cr& 
\sum_A\CO_A^2(0) \sum_{C,D}\CO_C \CO_D(x) \simeq {4\over x^d} \sum_{A,B}\CO_A \CO_B(0)~,\cr&
\sum_{A,B}\CO_A \CO_B(0) \sum_{C,D}\CO_C\CO_D(x) \simeq {4n\over x^d} \sum_{A,B}\CO_A\CO_B(0)~.
 }}
Combining these, and generalizing the analysis of subsection 3.1, we find the coupled beta functions 
\eqn\betadt{\eqalign{& {d\lambda\over d\log(\mu)}=\gamma(d) \lambda^2~,\cr&
\half {dc^2\over d\log(\mu)}=\gamma(d)c^2\lambda-\gamma(d)n{c^4\over 2}~.}}

The physical disordered theory is obtained for $n\rightarrow 0$ and we get 
\eqn\betadtf{\eqalign{& {d\lambda\over d\log(\mu)}= \gamma(d)\lambda^2~,\cr&
{dc^2\over d\log(\mu)}=2\gamma(d) c^2\lambda~.}}
If we take $\lambda=0$ initially, then $\lambda$ remains zero and disorder is exactly marginal, as we have already seen in subsection~3.1.
For any $\lambda>0$, $\lambda$ is marginally irrelevant and goes like $\gamma(d)\lambda\rightarrow -1/\log (\mu)$ in the infrared. Plugging this into the running of disorder, we see that disorder behaves like 
\eqn\rundis{c^2\rightarrow {1\over \log^2(\mu)}}
in the infrared. Disorder is thus marginally irrelevant in the present situation.

In order to make contact with the previous subsection, let the operator $\CO$ now have dimension $\Delta=\half d -\half\epsilon$ in the ultraviolet ($\epsilon>0$). The beta functions~\betadtf\ are then modified as 
\eqn\betadtfi{\eqalign{& {d\lambda\over d\log(\mu)}= -\epsilon \lambda+\gamma(d)\lambda^2~,\cr&
{dc^2\over d\log(\mu)}=-\epsilon c^2+2\gamma(d) c^2\lambda~.}} The coupling  $\lambda$ then evolves to its fixed point $\gamma(d)\lambda_*=\epsilon$, as for $c^2=0$, and the beta function for disorder becomes ${dc^2\over d\log(\mu)}=\epsilon c^2$. This is equivalent to the physics that we found in the previous subsection: disorder is irrelevant in the infrared according to the Harris bound, since the dimension of $\CO$ becomes $d-\Delta$ at low energies, and thus $c^2$ acquires effectively a negative mass dimension $\epsilon$.

\newsec{Holographic Analysis of Disorder}

In the AdS/CFT correspondence~\refs{\Mal,\Gubs,\Wit}, the scalar operator $\CO$ in the $d$-dimensional CFT maps to a scalar field $\phi$ in a gravitational theory on AdS$_{d+1}$. This field is weakly coupled in the large-$N$ limit of the CFT.
In the AdS/CFT correspondence it is natural to normalize the AdS action to have a factor of $c_T$ in front of it, so that otherwise
this action remains fixed in the large-$N$ limit. In this convention the two-point function $\vev{\CO \CO}$ scales as $c_T$, and
thus the source scales as $1/\sqrt{c_T}$ and disorder naturally scales as $1/c_T$. We will denote the source appearing in the
AdS equations of motion as ${\hat h}$ and denote its disorder averages by ${\hat c}^2$, noting that compared to our CFT
conventions in the previous sections ${\hat c}^2 \simeq c^2 / c_T$. On general grounds, because connected correlation functions
are suppressed by extra powers of $c_T$, the corrections from disorder to correlation functions that do not include the operator $\CO$ itself
will be suppressed by a power of $c_T$. In particular, this is true for the two-point function of the energy-momentum tensor, which will only be corrected at order one (compared to the leading term of order $c_T$), and it is also true for partition functions on spheres, that were computed in \Taka\ (following \GubserVV).

As in previous sections, we discuss Euclidean CFTs.\foot{See footnote 1 for a clarification.}  We choose a random  source for an operator ${\cal O}(x)$ of dimension~$\Delta$
\eqn\disorderx{
\overline{ \hh(x)} =0, \ \ \overline{ \hh(x) \hh(y) } =\hc^2 \delta^{(d)}(x-y).
}
In momentum space we have 
\eqn\disorderk{
\overline{\hh(k_1) \hh(k_2)} = \hc^2 \delta^{(d)}(k_1 + k_2), \qquad \qquad \qquad  |k_1| < k_{max},
}
where we explicitly put in a UV cutoff $k_{max}$ to regulate various divergences that we will find.
Obviously such a cutoff exists in any physical application.
We will be interested in particular in how the couplings and the disorder depend on this
cutoff. 
Our starting point is very similar to that of, for example,~\HartnollCUA, but with Euclidean disorder (which preserves the complete Poincar\'e group $ISO(d)$) instead
of Lorentzian disorder which is independent of time.\foot{One can also study holographic disorder using the methods of the holographic renormalization group \refs{\HeemskerkHK,\FaulknerJY}; this was done for electric disorder in \refs{\AdamsRJ,\AdamsYI}.}

We begin our analysis with free fields in AdS, corresponding to generalized free fields in the CFT (or to infinite $c_T$). We then add interactions in AdS, so that we include finite-$c_T$ effects. In principle in holography we should always include (for finite $c_T$) the coupling of the field $\phi$ to gravity, but for simplicity we will  work in the limit in which the $\phi^3$ and $\phi^4$ interactions of $\phi$ with itself are much larger than its interactions with gravity or any other interactions, and ignore the latter interactions. Gravitational and other interactions can be included by similar methods.

Note that because in the AdS/CFT correspondence~\refs{\Mal,\Gubs,\Wit} the
classical bulk action is the logarithm of the CFT partition function (at leading order in $1/N$),
in the holographic description we can compute the connected correlators for quenched random disorder
(such as \pathn) simply by putting random sources into the classical equations of motion and
averaging over them with a Gaussian weight. {\it Therefore, the AdS/CFT correspondence maps the problem of disorder in strongly coupled large-$N$ conformal field theories to standard stochastic differential equations.}

Note that at linear order
in the equations of motion of any field, it does not matter whether we
first solve its equations or first average over the disorder; at
higher orders this does matter, and we should first solve the equations
of motion for general sources and only then average over the disorder.

We will take the metric on AdS space to be
\eqn\adsmetric{
ds^2={{L^2}\over{z^2}}(dz^2+\delta_{mn}dx^m dx^n)~.
}
The boundary of space is at $z=0$, while large $z$ describes the IR region.
In our approximation where we neglect the coupling to gravity, this will remain the exact metric.
For a free scalar of mass $m$ on AdS the dimension of the dual operator is $\Delta={d\over 2} - \nu$, where
\eqn\nudef{
|\nu|=\sqrt{{{d^2}\over{4}}+m^2L^2}~.
}
When $\nu < 0$ ($\Delta > {d\over 2}$) the random source is an irrelevant deformation. In order to introduce a relevant deformation we have to use the alternative quantization for scalars in AdS space so that $\nu > 0$ \KlebanovTB, consistent with unitarity for operators of dimension ${d\over 2}-1< \Delta < {d\over 2}$. Note that the unitarity condition imposes $\nu<1$. 

On dimensional grounds we expect that if there is any part of the solution
that is independent of the cutoffs then it should depend only on $\hc z^{\nu}$ (after averaging
over the disorder). The perturbative
expansion in $\hc$ will thus break down (or acquire a cutoff
dependence) for large $z$ when $\nu > 0$, and for small $z$ when
$\nu < 0$, as expected from the (ir)relevance of the disorder.

The linear solution which is regular in the IR for an operator of dimension $\Delta$ with a  source $\hh(k)$ in momentum space is
\eqn\freesol{
\phi(x,z)= \int {d^d k\over {(2\pi)^{d/2}} } e^{i k \cdot x} \hh(k) z^{d/2} |k|^{-\nu} K_\nu(|k| z)~,
}
where $K_\nu$ is a modified Bessel function, up to  a constant ${{2^{1+\nu}}\over {\Gamma(-\nu)}}$  for $\nu \neq 0$, and $(-1)$
for $\nu=0$, which we will ignore.

Near the boundary, for $z \ll 1/k_{max}$, we expect the solution to be
governed by the source at the corresponding point,
so $\phi(x,z)$ should go as $\hh(x) z^{d-\Delta} = \hh(x) z^{d/2+\nu}$,
and this indeed follows from the behavior of the Bessel function at small
arguments. More precisely, this is true for $\nu < 0$, while for $\nu > 0$
the solution is dominated by the vacuum expectation value (VEV) of the corresponding operator so
it scales as $z^{d/2-|\nu|}$.
For the special case of $\nu=0$ the leading solution \freesol\ in this regime goes as
$\hh(x) z^{d/2} \log(z k_{max})$, again as expected from the form of the
non-normalizable solution for the scalar field.
As we go into the bulk the source gets smeared.

In the absence of interactions, the
scalar field $\phi$ vanishes after averaging over the disorder, so
let us compute the disorder average of $\phi^2$:
\eqn\freeaverage{
\overline{\phi^2(x,z)} = \hc^2 \int_0^{k_{max}} {{d^d k}\over {(2\pi)^d}}   |k|^{-2\nu} z^d K_\nu(|k| z)^2.
}
If we change variables $k_\mu = u_\mu/z$, then the average becomes
\eqn\vevs{
\overline{\phi^2(x,z)} =  z^{2\nu} \hc^2 \int_0^{z k_{max}} {{d^d u}\over {(2\pi)^d}}   |u|^{-2\nu}K_\nu(|u|)^2.
}

The integral is finite for $d>4\nu$, or $\Delta > d/4$. This coincides with the field theory result described after~\irgffvev, though we are not computing the same observable here. For $d=4\nu$  there are logarithmic divergences from the $|u|\to 0$ limit. 
For $\nu <1/2$ we have finite results for $d\geq 2$.
For $z \ll 1/k_{max}$ the average of $\phi^2$ goes as $\hc^2 z^{2\nu} (z k_{max})^{d-4\nu}$
when $\nu>0$, as $\hc^2 z^{2\nu} (z k_{max})^d$ for $\nu < 0$, and as $\hc^2 (z k_{max})^d \log^2(z k_{max})$ for $\nu=0$. In this
region the scalings are just the squares of the scaling of $\phi$. In the region $z \gg 1/k_{max}$ which describes the disordered field theory, the average goes as $\hc^2 z^{2\nu}$ in
all cases, including $\nu=0$, independently of the cutoff.

If the field $\phi$ is free in the bulk, then this is the whole story, and in the marginal case $\nu=0$  the disorder deformation seems to be exactly marginal.
Since there is no dependence on the cutoff, it seems that the field theory stays disordered with the same coefficient at all relevant scales. This is expected since the large-$N$ beta function for the disorder \betagff\ vanishes in this case. 
However, $\overline{\phi^2}$ does not have any direct field theory interpretation, so in the rest of this section we will
compute physical observables to understand better the dependence on $\hc^2$.

\subsec{Generalized Free Fields and Double-Trace Operators}

Let us start from the case of a free field theory in the bulk, in which \freesol\ is the exact expression for
the field in the bulk. This case was also analyzed using the replica trick in \Taka, with similar results. Up to constants, the expansion of \freesol\ in momentum space takes the form
(for $\Delta \neq d/2$)
\eqn\bdryexp{
\phi(k,z) = z^{d-\Delta} \hh(k) + z^{\Delta} \beta(k) + {\rm higher\ orders\ in\ }z,}
where $\beta(k) = |k|^{2\Delta-d} \hh(k)$ (up to constants mentioned above). $\beta(k)$ may be identified with the
VEV of the operator, so that we have
$\vev{{\cal O}(k)} = |k|^{2\Delta-d} \hh(k)$ for any source $\hh(k)$ (this is
determined by dimensional analysis; note that the vacuum expectation value here is a field
theory vacuum expectation value, not a disorder average). By taking a derivative
with respect to the source we obtain
\eqn\twopoint{
\vev{{\cal O}(k) {\cal O}(-k)} = |k|^{2\Delta-d}}
for any source $\hh(k)$. This is independent of the source, so obviously we obtain
the same answer also after the disorder averaging. General $n$-point functions of
${\cal O}(x)$ in this case are given by products of 2-point functions, so they are
all independent of the disorder; this trivially reproduces the field theory result~\connected\ (note that
the correlation functions we compute in this section are always connected).

In the special case of $\Delta=d/2$ we have instead 
\eqn\bdrymarginal{
\phi(x,z) = z^{d/2} (\hh(x) \log(z \mu) + \beta(x)),}
with a dependence of the VEV on the arbitrary scale $\mu$ which must be
introduced in this case. The form of the solution above implies that, up to a rescaling of $\mu$,\foot{By an abuse of notation, from now on we will use
$k$ both for the momentum vector and for its absolute value. We hope this will not cause any confusions.}
\eqn\vevmarginal{
\vev{{\cal O}(k)} = \beta(k) = \log(k/\mu) \hh(k)}
and
\eqn\twopointmarg{
\vev{{\cal O}(k) {\cal O}(-k)} = \log(k/\mu)~,}
independently of the disorder, as expected.

Next, let us add to the field theory action a double-trace interaction $S = {\hlambda \over 2}\int d^d x {\cal O}^2(x)$;
if we make $\hlambda$ space-dependent then this will also be useful for computing correlation functions of
${\cal O}^2(x)$, but for now let us just take $\hlambda$ to be a constant. Again, $\hlambda$ is related to
$\lambda$ in our CFT analysis by a factor of $c_T$. The analysis
will again be very similar to the field theory, so let us focus on the most interesting case of $\Delta=d/2$ where
we expect to see some logarithmic running~\betadtf. Since $\hlambda$ and $\hc^2$ run, we need to fix some UV cutoff $k_{max}$ as before, 
where we turn on the disorder with a value $\hc^2(k_{max}) = c_0^2$, and also set an initial boundary condition for the RG flow $\hlambda(k_{max}) = \lambda_0$.

In the presence of the double-trace deformation, \bdrymarginal\ is modified to~\WittenUA\
\eqn\phidouble{
\phi(x,z) = z^{d/2} ([\hlambda \beta(x) + \hh(x)] \log(z \mu) + \beta(x)) + {\rm\ higher\ orders\ in\ }z.
}
Using \vevmarginal\  for the new source in \phidouble, we have
\eqn\vevbeta{
\beta(k) = \log\left({k \over \mu}\right) (\hlambda \beta(k) + \hh(k)) \qquad \rightarrow \qquad \beta(k) = \vev{{\cal O}(k)} = {\hh(k) \log(k/\mu) \over {1 - \hlambda \log(k/\mu)}}.
}
This determines the 2-point function and all correlation functions of ${\cal O}(x)$, in agreement with the field
theory analysis of section 4. In particular, since the 2-point function is the same as the field theory result, and is independent
of the source, we find the same running of $\hlambda$ as we did in the field theory, namely that
$\hlambda(\mu) \propto -1 / \log(\mu)$.
Note that in general, $\hlambda$ could also start depending on the momentum due to higher derivative double-trace terms that could be generated, but we ignore this. 

If we now compute \phidouble\ at two different scales, one of them being the scale $k_{max}$ where we impose our initial boundary conditions, we obtain
(up to higher orders in $z$)
\eqn\scalerelation{
\eqalign{
\phi(x,z) & = z^{d/2} ([\hlambda(\mu) \beta(\mu, x) + \hh(\mu, x)] \log(z \mu) + \beta(\mu, x)) = \cr
& = z^{d/2} ([\lambda_0 \beta(k_{max}, x) + \hh(k_{max}, x)] \log(z k_{max}) + \beta(k_{max}, x)).
}}
Assuming that we take momenta much smaller than $\mu$, we can use \vevbeta\ to find
\eqn\scaleh{
\hh(\mu,k) = \hh(k_{max},k) {{1 - \hlambda(\mu) \log(k/\mu)}\over {1 - \lambda_0 \log(k/k_{max})}}.
}
Squaring this, and averaging over the disorder letting it depend on the scale and on the momentum (unlike the UV disorder $c_0^2$), we obtain
\eqn\scalec{
\hc^2(\mu,k) = c_0^2 \left( {{1 - \hlambda(\mu) \log(k/\mu)}\over {1 - \lambda_0 \log(k/k_{max})}} \right)^2.
}
We are interested in the $k$-independent part of this, corresponding to the Gaussian part of the disorder,
and indeed for $k \ll \mu < k_{max}$, the result is independent of $k$, and we find
\eqn\avgscalec{
\hc^2(\mu) = c_0^2 {{\hlambda^2(\mu)}\over {\lambda_0^2}}.
}
When joined with the running of $\hlambda$ that we found above, this implies
$\hc^2(\mu) \simeq 1 / \log^2(\mu)$, precisely as in the field theory analysis~\rundis.
Note that for $\hlambda=0$ we find no running of $\hc^2$, as expected.

In order to compute correlation functions of ${\cal O}^2(x)$ as in the previous section we need to take $\hlambda$ to depend on $x$
and to take derivatives with respect to it. The general principles of the AdS/CFT correspondence imply
that this must give the same answers as the field theory analysis above, but let us show this explicitly
for the one-point function of ${\cal O}^2(0)$ as a function of the source $\hh(x)$. To obtain this
we first compute $\vev{{\cal O}^2(0) {\cal O}(w)}$, which we can obtain by putting 
$\hlambda(x) = \lambda_1 \delta(x)$ (for simplicity we are not including a constant double-trace coupling here). 
With this coupling, \phidouble\ becomes in momentum space (at leading order in $z$)
\eqn\nphidouble{
\phi(k,z) = z^{d/2} \left(\left[\lambda_1 \int d^d l \beta(l) + \hh(k)\right] \log(z \mu) + \beta(k)\right).
}
Using the generalization of  \vevmarginal\ we obtain
\eqn\nvevmarginal{
\beta(k) = \log(k/\mu) \left[\lambda_1 \int d^d l \beta(l) + \hh(k)\right].}
Integrating this we obtain
\eqn\intnvevmarginal{
\int d^d k \beta(k) = \lambda_1 \int d^d k \log(k/\mu) \int d^d l \beta(l) + \int d^d k \log(k/\mu) \hh(k),}
leading to
\eqn\nintnvevmarginal{
\int d^d k \beta(k) = {{\int d^d k \log(k/\mu) \hh(k)} \over {1 - \lambda_1 \int d^d k \log(k/\mu)}}.}
Plugging this back into \nvevmarginal, we find
\eqn\noovev{
\vev{{\cal O}(k)} = \beta(k) =  \log(k/\mu) \left[\lambda_1  {{\int d^d l \log(l/\mu) \hh(l)} \over {1 - \lambda_1 \int d^d l \log(l/\mu)}} + \hh(k)\right].}
And, taking the derivative with respect to $\lambda_1$ and setting $\lambda_1=0$, we obtain
\eqn\ooovev{
\vev{{\cal O}(k) {\cal O}^2(x=0)} = \log(k/\mu) \int d^d l \log(l/\mu) \hh(l).}
This is also the derivative of $\vev{{\cal O}^2(x=0)}$ with respect to $\hh(k)$ (assuming we only keep
connected pieces), so we have
\eqn\oovev{\vev{{\cal O}^2(x=0)} = \left(\int d^d k \log(k/\mu) \hh(k)\right)^2.}
This is precisely the expected answer that we would get by contracting the two ${\cal O}$'s in this
operator with additional ${\cal O}$'s, so the results manifestly reproduce the field theory results
for generalized free fields. It is clear that this will be the case also for higher correlation functions,
and for theories with non-zero $\lambda$.

\subsec{Adding Bulk Interactions -- General}

Next, let us add interactions in the bulk. If we have some bulk potential (in addition to the mass term) 
$V(\phi) = \lambda_3 \phi^3 / 3 + \lambda_4 \phi^4 / 4$, 
then in momentum space (projecting the equation of motion of $\phi$ on a specific value of the momentum $k$) we have
\eqn\phiequation{
z^2 \partial_z^2 \phi + (1 - d) z \partial_z \phi - z^2 k^2 \phi - m^2 L^2 \phi = \lambda_3 \phi^2 + \lambda_4 \phi^3.
}
We will mostly specialize to the marginal case where $\Delta=d/2$ and $m^2 L^2 = -d^2/4$.

The general solution to the homogeneous equation in the marginal case is
\eqn\homo{
\phi(k,z) = z^{d/2} (A K_0(kz) + B I_0(kz))
}
for some constants $A$ and $B$, while the general solution to the inhomogeneous equation with a function $f(z)$ on the right-hand side of \phiequation\ 
(for a specific momentum $k$) is given by
\eqn\solution{
\phi(k,z) = 
z^{d/2} \left[ K_0(kz) \int_{z}^{\infty} {{ I_0(kw)}\over {w}} w^{-d/2} f(w) dw - (I_0 \leftrightarrow K_0) \right],
}
plus a homogeneous solution \homo. Assuming that both integrals converge when $z\to \infty$, and that the second integral vanishes in this limit, the IR boundary condition sets $B=0$. And, assuming that both integrals converge when $z\to 0$ (which will be the case for the functions $f$ that we will encounter), the fact that we do not want to have an extra source in the UV (which we could swallow into a redefinition of the original source) means that to get the precise solution we should choose $A$ such that the solution becomes
\eqn\solutionfinal{
\phi(k,z) = z^{d/2} \left[ -K_0(kz) \int_{0}^{z} {{ I_0(kw)}\over{w}} w^{-d/2} f(w) dw - I_0(kz) \int_{z}^{\infty} {{ K_0(kw)}\over{w}} w^{-d/2} f(w) dw \right].}
The new contribution to the vacuum expectation value of ${\cal O}$ is then given by
\eqn\vevsol{
\vev{{\cal O}(k)} = -\int_{0}^{\infty} {{K_0(kw)}\over{w}} w^{-d/2} f(w) dw.
}

For general values of the dimension of $\CO$, \solutionfinal\ takes exactly the same form but with the index of the Bessel functions modified from $0$ to $(-\nu)$, and \vevsol\ is replaced by
\eqn\vevsolnu{
\vev{{\cal O}(k)} = - k^{-\nu} \int_{0}^{\infty} {{K_{\nu}(kw)}\over{w}} w^{-d/2} f(w) dw.
}

\subsec{$\phi^4$ Interactions}

Let us start from the case $\lambda_3=0$. In this case the source in \phiequation\ starts at order $\hh^3$, and is given by 
\eqn\sourcethree{\eqalign{
f(k,z) = \lambda_4 \phi^3(k,z) =  \lambda_4 \int & {{d^d k_1}\over {(2\pi)^{d/2}} }  \hh(k_1) z^{3d/2} K_0(k_1 z) \cdot \cr & \cdot \int {{d^d k_2}\over {(2\pi)^{d/2}} } \hh(k_2) \hh(k-k_1-k_2) K_0(k_2 z) K_0(|k_1+k_2-k| z). \cr}}
The solution at order $\hh^3$ will be given by \solutionfinal.

If we are just interested in computing $\vev{{\cal O}(k)}$ at order $\hh^3$, and then using this to compute the disorder average of the two-point function $\vev{{\cal O}(k) {\cal O}(-k)}$ at order $\hc^2$, we can average already at this stage the disorder over two of the sources in  \sourcethree. Thus, up to this order we can replace \sourcethree\ by
\eqn\sourcethreen{\eqalign{
f(k,z) = \lambda_4 \phi^3(k,z) &= \hc^2 \lambda_4 \int_0^{k_{max}} d^d k_1 z^{3d/2} K_0(k_1 z)^2 \hh(k) K_0(k z)  \cr
 &= \hc^2 \lambda_4 \hh(k) K_0(k z) \int_{0}^{z k_{max}} d^d u_1 z^{d/2} K_0(u_1)^2}}
(up to a multiplicative constant). 

Plugging this into 
\solutionfinal, we find that the partly-averaged $\phi$ at order $\hh^3$, which we will denote by $\phi_{(3)}$, is given by
\eqn\solutionthreefour{\eqalign{
\phi_{(3)}(k,z) = \hc^2 \lambda_4 \hh(k) z^{d/2} & \left[ -K_0(kz) \int_{0}^{z} {{ I_0(kw)}\over {w}} \int_{0}^{w k_{max}} d^d u_1 K_0(u_1)^2  K_0(k w) dw - \right. \cr
& \left. \quad -I_0(kz) \int_{z}^{\infty} {{ K_0(k w)}\over {w}} \int_{0}^{w k_{max}} d^d u_1 K_0(u_1)^2  K_0(k w) dw \right]. \cr}}
Note that the integrals over $w$ indeed converge for finite cutoff, as is necessary for the validity of this specific solution.

Alternatively, we can compute the average vacuum expectation value $\overline{\langle{\cal O}(k) {\cal O}(-k)\rangle}$ directly by taking a derivative of \vevsol, and we find
\eqn\phifourvev{
\overline{\vev{{\cal O}(k) {\cal O}(-k)}} = - \hc^2 \lambda_4 \int_{0}^{\infty} {{K_0(k w)^2}\over {w}} \int_{0}^{w k_{max}} d^d u_1 K_0(u_1)^2  dw.}
Since the integral over $u_1$ converges as $k_{max} \to \infty$, it may seem that there is no cutoff dependence here (and thus also no interesting momentum dependence, given that the left-hand side is dimensionless). However, this is too naive. The integral over $w$ has two regions. When $w \gg 1/k_{max}$, the integral over $u_1$ is approximately given by the integral from $0$ to $\infty$, which is some constant $C_0$ independent of $w$. However, when $w \ll 1/k_{max}$, the integrand can be approximated by $\log^2(u_1)$, such that the integral over $u_1$ behaves as $(w k_{max})^d \log^2(w k_{max})$. Thus, we have approximately, for $k \ll k_{max}$ (up to constants)
\eqn\nphifourvev{
\eqalign{
\overline{\vev{{\cal O}(k) {\cal O}(-k)}} & \simeq - \hc^2 \lambda_4 \left[ \int_{0}^{1/k_{max}} {{K_0(k w)^2}\over{w}} \left( w k_{max} \right)^d \log^2 \left( w k_{max} \right)  dw \right. \cr &\left. \qquad\qquad\qquad +  \int_{1/k_{max}}^{\infty} C_0 {{K_0(kw)^2}\over{w}} dw \right] \cr
& \simeq -\hc^2 \lambda_4 {C_0\over 3} \log^3(k/k_{max}) + O\left(\hc^2 \lambda_4 \log^2(k/k_{max}) \right). \cr}}
Note that this result is dominated by the contribution from $w \gg 1/k_{max}$ which is above the effective short-distance cutoff on the radial direction.
We will try to understand this surprising result from the field theory point of view in the next section.

Repeating exactly the same procedure for general values of the operator dimension $\Delta < d/2$, \phifourvev\ is replaced by
\eqn\phifourvevnu{
\overline{\vev{{\cal O}(k) {\cal O}(-k)}} = - \hc^2 \lambda_4 k^{-2\nu} \int_{0}^{\infty} {{K_{\nu}(k w)^2}\over {w}} w^{2\nu} \int_{0}^{w k_{max}} d^d u_1 u_1^{-2\nu} K_\nu(u_1)^2  dw.}
Using the expansion of the Bessel function near $x=0$, $K_{\nu}(x) \propto x^{-\nu}$, we find that the contribution from the region $w > 1/k_{max}$ is now proportional to ${\hat c}^2 \lambda_4 k^{-4\nu} \log(k/k_{max})$, while the region $w < 1/k_{max}$ has a contribution proportional to $k^{-4\nu}$. Thus, in this case we have
\eqn\phifourvevnnu{
\overline{\vev{{\cal O}(k) {\cal O}(-k)}} \simeq - \hc^2 \lambda_4 C_{\nu} k^{-4\nu} \log(k/k_{max})}
for some constant $C_{\nu}$,
with an extra logarithm on top of the power expected on dimensional grounds.
Again, we will understand the origin of this logarithm in the next section.

\subsec{$\phi^3$ Interactions}

Now let us take a non-zero value of $\lambda_3$, so that the operator $\cal O$ has a non-zero 3-point function proportional to $\lambda_3$ (for simplicity we take $\lambda_4=0$ here). In this case we have a source for $\phi$ already at order $\hh^2$, given by
\eqn\sourcetwo{
f(z) = \lambda_3 \phi^2(k,z) =  \lambda_3 \int {{d^d k_1}\over {(2\pi)^{d/2}} }  \hh(k_1) z^{d} K_0(k_1 z) \hh(k-k_1) K_0(|k-k_1| z).
}
Performing a disorder average here gives something proportional to $\delta^{(d)}(k)$, related to a constant VEV for the disorder-averaged $\phi^2$ \vevs.

Anyway, to go to higher orders we now cannot directly average over this, and we have to plug this full expression into \solutionfinal, leading to (up to a multiplicative constant)
\eqn\solutiontwo{\eqalign{
\phi_{(2)}(k,z) & = \lambda_3 z^{d/2} \int d^d k_1 \hh(k_1) \hh(k-k_1) \cdot \cr
& 
\qquad\quad \left[ -K_0(kz) \int_{0}^{z} {{ I_0(kw)}\over {w}} w^{d/2} K_0(k_1 w) K_0(|k-k_1|w)dw - \right. \cr
& \left. \qquad\qquad I_0(kz) \int_{z}^{\infty} {{ K_0(kw)}\over{w}} w^{d/2} K_0(k_1 w) K_0(|k-k_1|w) dw \right].}}

At order $\hh^3$ the source is now given by this solution times the first order solution, namely
\eqn\soltwothree{\eqalign{
f(z) = \lambda_3 (\phi^2(k,z))_{(3)} & = 2 \lambda_3^2 z^d \int d^d k_2 \hh(k-k_2) K_0(|k-k_2| z) \int d^d k_1 \hh(k_1) \hh(k_2-k_1) \cdot \cr
& 
\qquad\quad \left[ -K_0(k_2z) \int_{0}^{z} {{ I_0(k_2w)}\over{w}} w^{d/2} K_0(k_1 w) K_0(|k_2-k_1|w)dw - \right. \cr
& \left. \qquad \qquad I_0(k_2z) \int_{z}^{\infty} {{ K_0(k_2w)}\over{w}} w^{d/2} K_0(k_1 w) K_0(|k_2-k_1|w) dw \right].}}

Once again, for the purpose of computing the disorder average of $\vev{{\cal O}(k) {\cal O}(-k)}$ to order $\hc^2$, we are allowed to average over two of the disorder coefficients appearing here. 
There is one average here that gives a divergence, which is when we average over the last two sources such that $k_2=0$. 
%We will identify the source of this divergence in the next section.
The origin of this divergence is confusing at first sight, since it does not go away even when we put both UV and IR
cutoffs on the disorder distribution \disorderk. However, if we follow its origin from the bulk point of view, we see
that it arises from an exchange of a bulk scalar with spatial momentum $k_2=0$. Thus, this is an IR divergence
that will disappear in the presence of any IR cutoff on the full theory. We will therefore subtract it in our computation.

Considering the other contributions to the average, we obtain
\eqn\phithree{\eqalign{
f(z) = \lambda_3 (\phi^2(k,z))_{(3)} & = 2 \hc^2 \lambda_3^2 z^d \hh(k) \int d^d k_1 K_0(k_1 z) \cdot \cr
& 
\qquad \left[ -K_0(|k+k_1| z) \int_{0}^{z} {{ I_0(|k+k_1| w)}\over{w}} w^{d/2} K_0(k_1 w) K_0(k w)dw - \right. \cr
& \qquad \quad  \left. I_0(|k+k_1| z) \int_{z}^{\infty} {{ K_0(|k+k_1| w)}\over{w}} w^{d/2} K_0(k_1 w) K_0(k w) dw \right].}}

Next we need to plug this into \solution\ to obtain the third order solution, but for us it is enough to obtain the VEV of ${\cal O}$ to third order, using \vevsol. Using this to compute the 2-point function, we obtain
\eqn\twotwon{\eqalign{
\overline{\vev{{{\cal O}(k) {\cal O}(-k)}}} = - 2 \hc^2 \lambda_3^2 & \int_0^{\infty} dv {{K_0(kv)}\over{v}} v^{d/2} 
\int_{0}^{k_{max}} d^d k_1 K_0(k_1 v) \cdot \cr
& 
\left[ -K_0(|k+k_1| v) \int_{0}^{v} {{ I_0(|k+k_1| w)}\over{w}} w^{d/2} K_0(k_1 w) K_0(k w)dw - \right. \cr
& \left. \quad I_0(|k+k_1| v) \int_{v}^{\infty} {{ K_0(|k+k_1| w)}\over{w}} w^{d/2} K_0(k_1 w) K_0(k w) dw \right].}}
The integrals here converge separately, but as in the previous subsection we need to analyze them more carefully, and such a careful analysis again gives a $\log^3(k/k_{max})$ behavior. 

Note that in this case, and also in the previous case at higher orders in $\hc^2$, it is important that we first
solve the equations for arbitrary sources, and only then average over the disorder. We would get different
answers if we would first average and then plug the results into the higher order computations. Thus, it
is not clear if there is any meaning to an effective disordered solution for the bulk field $\phi(x,z)$.

\newsec{$1/N$ Corrections}

In subsections 5.3 and 5.4 we saw how we could perform holographic computations for disorder with $c^2$ of order the
central charge $c_T$, at leading order in the disorder $\hc^2 = c^2 / c_T$ (in principle it is straightforward to extend this computation
to arbitrary orders). We found several interesting results for this scaling of $c^2$, and in this section
we would like to understand these results from the field theory point of view. So, we consider the leading
correction as a function of $c^2/c_T$ directly in the field theory. For finite $c^2$ in the large-$N$ limit this
gives the first $1/N$ correction, that becomes important (in the marginal case) at distances exponentially
large in $N$, while for $c^2$ of order $c_T$ this correction is significant at distances of order one, and can lead to changes even
if $c^2/c_T$ is small compared to other scales. Below we will present two derivations of our results in parallel. One approach is direct, and the other is through the replica trick. It seems like the latter approach is easier to generalize to higher orders, while the direct approach removes doubts about the validity of the replica trick in this situation.

In principle it is possible to compute $\overline{\vev{\CO \CO}}$ in a perturbative expansion in the disorder, using
conformal perturbation theory. At leading order this formally gives
\eqn\pertlead{
\overline{\vev{\CO(x) \CO(y)}} = {1 \over (x-y)^{2\Delta}} +
                          {1\over 2} c^2 \int d^d w \vev{\CO(x) \CO(y) \CO(w) \CO(w)} + \cdots.}
The limit when the two points near $w$ come together is singular, but we can introduce some UV cutoff on the distance between the two points and use the
OPE expansion in the original theory near $w$, which has the general schematic form
\eqn\newOPE{
\CO(w) \CO(w+\epsilon) \sim \sum_P c_{OOP} P(w) \epsilon^{\Delta_P-2\Delta}.}
The identity operator can be dropped from~\newOPE\ since it gives a
disconnected contribution already for fixed sources. We will normalize the
operators $P$ to have a 2-point function with coefficient one. 
Terms with $\Delta_P < 2\Delta$ are UV-singular, and will depend on a UV cutoff $\mu$
on the distance $1/\epsilon$. 
Terms
with $\Delta_P \simeq 2\Delta$ lead to universal contributions and should be kept, while terms with
$\Delta_P > 2\Delta$ will be negligible as the cutoff is removed.  As we are studying large-$N$ theories, we will assume that an operator with $\Delta_P \simeq 2\Delta$ exists. We denote $P=\CO^2$ for this particular operator (keeping in mind that at finite but large $N$ the dimension is no longer exactly additive, $\Delta_{\CO^2}-2\Delta_O\sim c_T^{-1}$). There could be other operators of dimension $2\Delta$, in particular, single-trace operators. These would modify the results below, but it is clear how to generalize our discussion when necessary.

The contribution from $\CO^2$ at the order $c^2$ is
\eqn\pertope{
\eqalign{
\overline{\vev{\CO(x) \CO(y)}} &={1\over2} c^2 c_{OO\CO^2}  \int d^d w\vev{\CO(x) \CO(y) \CO^2(w)} \mu^{2\Delta-\Delta_\CO^2} = \cr
               &= {1\over2} c^2c_{OO\CO^2}^2   \mu^{2\Delta-\Delta_{\CO^2}} \int d^d w {1\over {(x-y)^{2\Delta-\Delta_{\CO^2}} (x-w)^{\Delta_{\CO^2}} (y-w)^{\Delta_{\CO^2}}}}.}}

The above result arises in the replica trick as follows. We study $n$ copies of the CFT, perturbed by $\half c^2\sum_{A,B} \CO_A\CO_B$. The terms with $A=B$ are interpreted as the operators $\mu^{2\Delta-\Delta_\CO^2}\CO^2_A$. To compute the average~\pertope\ we then consider the correlator $\langle\CO_1\CO_1\rangle$ and in the end of the computation we take $n\rightarrow 0$ (without taking a derivative with respect to $n$).
Clearly the only contribution at order $c^2$ arises from $\int d^dw\langle\CO_1(x) \CO_1^2(w)\CO_1(y)\rangle$, and it is independent of $n$. Therefore, one finds precisely~\pertope.

The integral over $w$~\pertope\ diverges at large $w$ if $\Delta_{\CO^2} \leq d/2$.
In the large-$N$ limit this happens when $\Delta\leq d/4$, giving another exposition of the problems
in that case (that we have already discussed at length in previous sections). 

As long as ${d\over 2}<\Delta_{\CO^2} < d$ the integral converges both for large $w$
and for $w\to x$ and $w\to y$. We get (up to constants) the following leading order in $c^2$ result 
\eqn\pcont{
\overline{\vev{\CO(x) \CO(y)}}
          = {1\over 2} c^2  \mu^{2\Delta-\Delta_{\CO^2}} {c_{OO\CO^2}^2 \over  {(x-y)^{2\Delta+\Delta_{\CO^2}-d}}}.}
This result is to be trusted only for
$\Delta_{\CO^2}$ equal to $2\Delta$ up to infinitesimally small
corrections. It is easy to see both in the direct approach and in the replica trick, that in the absence of such an operator there is no contribution at order $c^2$ that is independent of the cutoff.

If the dimension $\Delta_{\CO^2}$ is exactly $2\Delta$, \pcont\ 
just gives a finite contribution going as
$c^2 c_{OO\CO^2}^2 / (x-y)^{4\Delta-d}$.
In the large-$N$ limit, when we normalize the 2-point function of
$\CO^2$ to one, 
$c_{OO\CO^2}$ behaves at large-$N$ as $\sqrt{2}(1 + A / c_T + ...)$
for some constant $A$ independent of $c_T$.
For $c^2\sim 1$ only
the first term remains, while for $c^2\sim c_T$ the second term also
remains. If $c^2<<c_T$ but $c^2/c_T$ stays constant in the large $N$ limit, then we can trust our approximation and neglect higher order corrections in $c$ and higher order corrections in $c_T^{-1}$.

We are interested in comparing with the holographic computation which
sees only the disorder-averaged free energy, so we should subtract
from \pcont\ the disconnected piece, which is always present and goes (at order $c^2$) as
\eqn\discont{
\overline{\vev{\CO(x)}\vev{\CO(y)}}
          = \int d^d w {c^2 \over  {(x-w)^{2\Delta} (y-w)^{2\Delta}}}.}
At leading order in $1/c_T$ this exactly cancels the leading term above, as in our general
discussion of generalized free field theories in section 4.\foot{This cancellation is very easy to see in terms of the replica trick. The connected averaged correlation function corresponds to the derivative at $n=0$ of  $\langle \sum_A\CO_A\sum_B\CO_B \rangle$, which goes like $n^2c^2$ in generalized free field theory. This is why the connected averaged correlation functions of the single-trace operators do not depend on disorder in a generalized free field theory.} All
remaining terms are of order $c^2/c_T$, ensuring a finite limit
when $c^2 \propto c_T$ with $c^2/c_T\ll 1$.

Thus, for $\Delta_{\CO^2} = 2\Delta < d$,
the connected two-point function at order $c^2$ goes as
\eqn\opecont{
{A c^2 \over {c_T (x-y)^{4\Delta-d}}}.}

We will now include the effects that are associated to $\Delta_{\CO^2}$ deviating from $2\Delta$. In general
$\Delta_{\CO^2} = 2\Delta - B / c_T + \cdots$ for some constant $B$, which stays finite as $c_T\rightarrow\infty$.
In such a case we can write
$\mu^{2\Delta-\Delta_P} = 1 + (B / c_T) \log(\mu) + \cdots$.
The term of order one contributes to our computation a term proportional to $A$
as in \opecont. The $B$-term gives a contribution to the 2-point function \pcont\ proportional to
\eqn\bcont{
{c^2 B \log(\mu |x-y|) \over {c_T (x-y)^{4\Delta-d}}}.}
Note that in the holographic computation $B$ should have contributions 
proportional both to
$\lambda_4$ and to $\lambda_3^2$, and in particular we found
a contribution of this form proportional to $\lambda_4$ in our computation with
general dimensions $\Delta$ \phifourvevnnu. 

Our computation up to now was done for the relevant case, $\Delta < d/2$.
In the marginal case $\Delta=d/2$ we need to reconsider the integral
over $w$ that we did above, since it has a logarithmic UV divergence as $w\to x$
and $w\to y$. Note that this divergence has a different origin from the one
leading to \bcont\ -- it is related to the fact that $\Delta_{\CO^2}$ is close to $d$  (but it is still also close to $2\Delta$). Note also that the
disconnected contribution \discont\ has a similar logarithmic divergence
whenever $\Delta$ is close to $d/2$. 
We therefore need to combine all these logarithms carefully.

In a general large-$N$ theory, the dimension of the single-trace operator ${\cal O}$ in the large-$N$ limit goes as
$\Delta = d/2 - D / c_T + \cdots$ for some constant $D$. 
If we now expand \discont\ in $1/c_T$ we get (keeping only the leading logarithm)
\eqn\discontn{\eqalign{
\overline{\vev{\CO(x)}\vev{\CO(y)}}
          &= \int d^d w {c^2 \over  {(x-w)^{d} (y-w)^{d}}} \left(1 + (d-2\Delta) (\log(\mu(x-w)) + \log(\mu(y-w))) + \cdots \right) = \cr
&= 3(d-2\Delta){\gamma(d) c^2 \over (x-y)^d}  \log^2(\mu(x-y)) + O\left({1\over c_T^2} \right),}
}
dropping the leading term (which eventually cancels) in the second line.
%where in the last line we wrote the most singular terms up to order $c^2/c_T$.
Similarly, the integral over $w$ appearing in the second line of \pertope\ for $\Delta_{\CO^2} \simeq d$
gives (again keeping only the leading logarithm)
\eqn\pertopen{
3(d-\Delta_{\CO^2}){\gamma(d)\over (x-y)^d}  \log^2(\mu(x-y)) + O\left({1\over c_T^2}\right)~,}
so that the total contribution to $\overline{\vev{\CO(x) \CO(y)}}$ goes as
\eqn\pertopenn{(3 d - 3 \Delta_{\CO^2} - 2 (2 \Delta - \Delta_{\CO^2})) {\gamma(d) c^2  \over (x-y)^d} \log^2(\mu(x-y)) + O\left({1\over c_T^2}\right)~
.}
When we subtract \discontn\ from \pertopenn\ to obtain the connected two-point function, the contribution is of order $c^2/c_T$ and the leading log term takes the form
\eqn\twopointconn{(2 \Delta - \Delta_{\CO^2}) {\gamma(d) c^2 \log^2(\mu(x-y)) \over (x-y)^d}.}

This is exactly the same form that we found in the previous section, since the Fourier transform
of  $\log^2(\mu(x-y)) / (x-y)^d$ is proportional to $\log^3(k/\mu)$. So we identify the coefficient 
we found there as the correction at order $1/c_T$ to the dimension of $\CO^2$ compared to
twice the dimension of $\CO$. It is indeed well-known from the early days of the AdS/CFT correspondence
\wittenunpub\ that the diagrams that appeared in the
previous section (at order $\lambda_4$ and at order $\lambda_3^2$) contribute to this `anomalous dimension'.\foot{Note
that in the previous subsection we neglected any contributions from other bulk fields to the correlators we computed, such
as graviton exchange. However,
we expect the leading logarithmic behavior to be the same for all of these contributions, corresponding to different contributions to the
anomalous dimension for $\CO^2$, and this is confirmed by our analysis in this subsection.}
As in previous cases, the two-point function of the form \twopointconn\ does not look like
an anomalous dimension for $\CO$ in the disordered theory. So the interpretation of the computations above in terms of anomalous dimensions is not straightforward. 
To learn about the low-energy behavior we need to go to higher orders, and to resum the perturbative expansion in $c^2/c_T$. This is beyond the
scope of this paper, but it would be very nice to do so in the future.

\bigskip
\noindent{\bf Acknowledgments}

We would like to thank Amnon Aharony, John Cardy, Guy Gur-Ari, Sean Hartnoll, Carlos Hoyos, Elias Kiritsis, Slava Rychkov, Jorge Santos, David Simmons-Duffin, and Tadashi Takayanagi for useful discussions. This work was supported in part by an Israel Science Foundation center for excellence grant
and by the I-CORE program of the Planning and Budgeting Committee and the Israel Science Foundation (grant number 1937/12). The work of OA was also supported in part by the Minerva foundation with funding from the Federal German Ministry for Education and Research, by a Henri Gutwirth award from the Henri Gutwirth Fund for the Promotion of Research, and by the ISF within the ISF-UGC joint research program framework (grant no. 1200/14). OA is the Samuel Sebba Professorial Chair of Pure and Applied Physics. ZK is also supported by the ERC STG grant 335182 and by the United States-Israel Bi-national Science Foundation (BSF) under grant 2010/629. The work of SY was also supported in part by BSF (grant 2012/383) and GIF (grant I-244-303.7-2013).

\listrefs

\end